\documentclass[journal]{IEEEtran}
\usepackage{amsmath,amsfonts}
\usepackage{graphicx}

\usepackage{dcolumn}
\newcolumntype{d}[1]{D{.}{.}{#1}}
\usepackage{multirow}
\usepackage{amssymb}
\usepackage{bm}  

\renewcommand{\vec}[1]{\text{\boldmath$#1$}}

\begin{document}

\title{Theory of the linewidth--power product of photonic--crystal surface--emitting lasers}

\author{Hans Wenzel, Eduard Kuhn, Ben King, Paul Crump,~\IEEEmembership{Senior Member,~IEEE}, and Mindaugas Radziunas
\thanks{Hans Wenzel, Ben King, and Paul Crump are with the Ferdinand--Braun--Institut (FBH), Gustav-Kirchhoff-Str. 4, 12489 Berlin, Germany.}
\thanks{Eduard Kuhn and Mindaugas Radziunas are with the Weierstrass Institute
for Applied Analysis and Stochastic (WIAS), Mohrenstr. 39, 10117 Berlin, Germany.
}
\thanks{
This work has been submitted to the IEEE for possible publication. 
Copyright may be transferred without notice, after which this version may no longer be accessible.
}}

\markboth{\MakeLowercase{submitted to} IEEE Journal of Quantum Electronics}%
{Wenzel \MakeLowercase{\textit{et al.}}: The linewidth--power product of photonic--crystal surface--emitting lasers}

\IEEEpubid{0000--0000/00\$00.00~\copyright~2024 IEEE}

\maketitle

\begin{abstract}
A general theory for the intrinsic (Lorentzian) linewidth of photonic--crystal surface--emitting lasers (PCSELs) is presented. 
The effect of spontaneous emission is modeled by  a classical Langevin force entering the equation for the slowly varying waves.
The solution of the coupled--wave equations, describing the propagation of four basic waves within the plane of the photonic crystal, is  expanded in terms of the solutions of the associated spectral problem, i.e. the laser modes.
Expressions are given for photon number, rate of spontaneous emission into the laser mode, Petermann factor and effective Henry factor entering the general formula for the linewidth.
The theoretical framework is applied to the calculation of the linewidth--power product of air--hole and all--semiconductor PCSELs.
For output powers in the Watt range, intrinsic linewidths in the kHz range are obtained in agreement with recent experimental results.
\end{abstract}


\section{Introduction}
\IEEEPARstart{P}{hotonic--}crystal surface--emitting lasers (PCSELs) belong to a new class of diode laser distinguished from edge--emitting lasers (EELs) and vertical--cavity surface--emitting lasers (VCSELs) \cite{noda2017photonic}.
In all of these lasers lasing is achieved in the same way by applying an electrical bias between Ohmic contacts deposited at the bottom and top surfaces. 
Electrons and holes are injected from the n-- and p--doped layers into an active layer where they recombine.

In EELs the optical waves generated by stimulated recombination travel parallel to the epitaxial layers along a preferred axis and are fed back by the facets normal to, or by Bragg gratings along, the propagation axis. The light is coupled out of the facets. Typically the emitted beam is elliptic and highly divergent along the fast axis (perpendicular to the layers). 
Single--transverse--mode continuous-wave (cw) output powers from single emitters are in the Watt range \cite{epperlein2013semiconductor}. Internal wavelength-stabilization can be realized either in a distributed--feedback (DFB) configuration or by a distributed Bragg reflector (DBR).

In VCSELs the waves propagate perpendicular to the layers and are fed back by DBRs.
The light is coupled out from the wafer surface. 
The emitted beam can be circular but is still divergent. 
The single--mode output powers are in the mW range \cite{michalzik2012vcsel}. 

\begin{figure}[!t]
\centering
\includegraphics[width=3in]{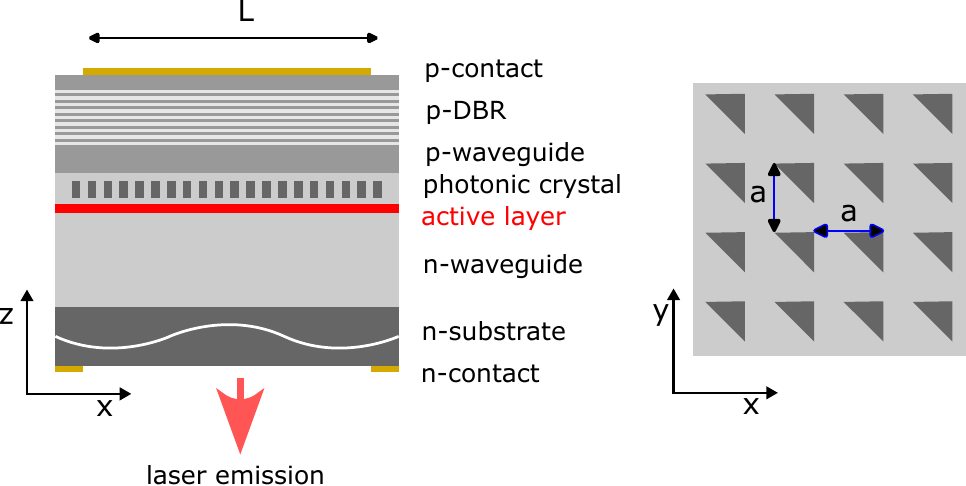}
\caption{Cross--sectional view of a PCSEL (left) and top view of the PC layer (right).}
\label{fig:PCSEL_schematic}
\end{figure}

A PCSEL shown schematically in Fig.~\ref{fig:PCSEL_schematic} can be considered to be in some sense a combination of an EEL and a VCSEL.
As in an EEL, the optical waves generated by stimulated recombination travel parallel to the layers, but into all in--plane directions. Feedback is provided by a two--dimensional photonic crystal (PC) in a plane that is offset near to  the active layer with a periodicity corresponding to second order Bragg diffraction for wavelengths around the peak of the gain spectrum. 
Therefore, a part of the optical field is radiated vertically. 
To force the out--coupling from one surface, a single DBR like in a VCSEL can be implemented, too.
A properly designed PCSEL can emit  tens of Watt of optical power in a circular beam with a very small divergence \cite{yoshida2023high}.

\IEEEpubidadjcol

Due to the fact that a PCSEL combines large emitting area, single--transverse mode operation, integrated wavelength stabilization, and high output power a small spectral linewidth could be expected. 
Thus, PCSELs emitting at $1064$~nm, for example, could replace optically pumped Nd:YAG non--planar ring oscillators (NPROs) used for coherent optical communication in space \cite{schwander2018lctsx}.

Experimental results published  in the past revealed intrinsic (Lorentzian) linewidths around $70$~kHz at output powers of several tens of mW \cite{kalapala2021linewidth, saleeb2022spectral, inoue2023measurement}. 
Recently, a spectral linewidth as low as $1.23$~kHz at an output power of $3$~W was reported \cite{noda2023high}.

The calculation of the spectral linewidth of PCSELs was based until now on the numerical solution of the time--dependent coupled--wave equations taking into account several sources of noise leading to a finite linewidth \cite{inoue2023measurement}.
Note, however, that for a spectral resolution of $1$~kHz a time trace of $1$~ms has to be simulated. 
Such simulations do not allow big parameter sweeps because of the large computational time, albeit it could be reduced to some extent by zero padding in the discrete Fourier transformation and using large spatial and temporal simulation steps \cite{chan1997semiconductor}.
The aim of this paper is to derive an expression for the spectral linewidth of PCSELs based on an expansion of the solution of the coupled--wave equations into the modes and performing a single-mode approximation. 
The modes can be  obtained directly as solutions of the coupled--wave equations in the frequency domain (i.e. solutions of the so-called spectral problem) at threshold or by solving the time--dependent equations together with dynamical equations for the carrier density to determine  the steady--state above threshold as done in \cite{Wenzel2021}. 
The corresponding time trace to be simulated amounts then only several ns (excluding turn-on behavior).

Spontaneous emission sets the ultimate limit of the  intrinsic linewidth, \emph{i.e.} the full width at half maximum of the Lorentzian shape of the power spectral density (PSD) of the optical field corresponding to a flat PSD of the optical frequency fluctuations.
Although a quantum phenomenon, spontaneous emission is treated here by a classical Langevin noise source as appropriate for the calculation of the spectral linewidth \cite{tromborg1994traveling}.
Other noise sources due to fluctuations of the charged carriers, injection current, and temperature, partially occurring on different time scales resulting in $1/f$ noise and a Gaussian shape of the field PSD are not considered \cite{mercer1991}. 
Moreover, a single--mode approximation is employed neglecting the impact of side modes.

There are several methods to simulate the optical fields of PCSELs. Among them are the plane wave expansion (PWE) method, 
the rigorous coupled--wave analysis (RCWA), the finite--difference time--domain (FDTD) method, and the coupled--wave theory (CWT).
A detailed comparison of the pros and cons of the different methods can be found in Ref. \cite{noda2023high}. 
The PWE method \cite{sukhoivanov2009} and RCWA \cite{song2018} work both in the frequency domain and assume an infinite in-plane periodic structure. 
Additionally, the PWE employed to PCSELs has to assume an infinite thickness of the photonic crystal. 
The FDTD method \cite{sukhoivanov2009} working in the time domain can handle both infinite and finite in--plane structures as well as finite thicknesses. However, for the simulation of PCSELs with a large size FDTD is not feasible in terms of computation time and is not suited for simulations above laser threshold.
In contrast to the previous methods, which solve Maxwell's equations numerically exact in two or three dimensions, CWT is a semi--analytic method based on a cleverly chosen Ansatz for the optical fields. The CWT, previously derived in a couple of papers \cite{Liang2011,Peng2011,Liang2012,Inoue2019} based on earlier work on DFB lasers \cite{Shams2000,streifer1975coupling},
allows a quasi--three dimensional simulation of the optical fields of finite--sized PCSELs in frequency and time domains as well as above threshold. In Ref. \cite{wang2017}, CWT and FDTD method were compared and a mutual good agreement was found.

The content of the paper is as follows. First, the basic equations for the computation of the optical field taking into account dispersion and spontaneous emission are presented. Second, the time--dependent coupled--wave equations are summarized. Third, the spectral problem is considered and a number of useful  characteristics are presented. Fourth, an expression for the spectral linewidth is derived based on the model presented in \cite{Wenzel2021} and an orthogonality relation for the solutions of the spectral problem (the laser modes). Finally, numerical results including the linewidth power--product are given for an air--hole PCSEL and an all--semiconductor PCSEL.

\section{Basic equations}

Assuming a nearly harmonic dependence on the time $t$ with a reference frequency $\omega_0$ (or vacuum wavelength $\lambda_0$), the real--valued electric field strength and polarization of the medium are given by
\begin{equation}
\label{eq:Ansatz_time}
\vec{\mathcal{E}}(\vec{r},t)=\frac{1}{2}\vec{E}(\vec{r},t)e^{i\omega_0t}+\text{c.c.}
\end{equation}
and
\begin{equation} 
\begin{aligned}
\label{eq:Peq}
\vec{\mathcal{P}}(\vec{r},t) 
&= \varepsilon_0 \int_0^\infty \hat{\chi}(\vec{r},\tau) \vec{\mathcal{E}}(\vec{r}, t-\tau) \, d\tau \\
&\approx\frac{\varepsilon_0}{2} \int_0^\infty \hat{\chi}(\vec{r},\tau)\Big[\vec{E}(\vec{r},t) \\
&\hspace{7mm} - \tau\partial_t \vec{E}(\vec{r},t)\Big]e^{-i\omega_0\tau} \, d\tau e^{i\omega_0t}+\text{c.c.}\\
&=\frac{\varepsilon_0}{2} \Big[\chi(\vec{r},\omega_0) \vec{E}(\vec{r},t) \\
&\hspace{5mm} - i\partial_\omega
\chi(\vec{r},\omega)|_{\omega = \omega_0} \partial_t \vec{E}(\vec{r},t)\Big]e^{i\omega_0t}+\text{c.c.}
\end{aligned},
\end{equation}
respectively, where $\vec{r}=(x,y,z)$ is the vector of the spatial coordinates $x$, $y$ (in--plane) and $z$, c.c. denotes the complex conjugate, $i$ the imaginary unit, and $\partial_t$, $\partial_\omega$ the partial derivatives with respect to $t$, $\omega$, respectively.
The replacement of $\vec{E}(\vec{r}, t-\tau)$ by its first order approximation 
\linebreak $\vec{E}(\vec{r},t)-\partial_t \vec{E}(\vec{r},t) \tau$  bases on the assumptions that $\vec{E}(\vec{r},t)$ is slowly varying with respect to time and that the real--valued susceptibility $\hat{\chi}(\vec{r},\tau)$ decays quickly to zero with increasing $\tau$. 
The susceptibility depends  on temporally slowly varying variables such as excess carrier density $N$ in the active layer and temperature $T$. 
Note, that the restriction of the $\tau$--integration to the interval $[0,\infty]$ is a consequence of causality, which implies $\hat{\chi}(\mathbf{r},\tau)=0$ for $\tau<0$. 
 
After inserting \eqref{eq:Ansatz_time}, the corresponding expression for the magnetic field strength, and \eqref{eq:Peq} into Maxwell's equations and eliminating the magnetic field strength, the slowly temporally varying field $\vec{E}$ obeys the equation
\begin{multline}
\label{eq:Helmholtz}
-\vec{\nabla}\times\left(\vec{\nabla}\times\vec{E}(\vec{r},t)\right)+k_0^2\varepsilon(\vec{r},N,T)\vec{E}(\vec{r},t)\\=2ik_0\frac{n(\vec{r}) n_\mathrm{g}(\vec{r})}{c}\partial_t \vec{E}(\vec{r},t) +  \vec{D}_\mathrm{disp}(\vec{r},t) + \vec{F}_\mathrm{sp}(\vec{r},t)
\end{multline}
in basic agreement with \cite{berneker2010dynamische,Wenzel2013}.
Here,
$k_0=\omega_0/c=2\pi/\lambda_0$ is the wavevector of free space, $c$ is the vacuum speed of light,
\begin{equation}
\varepsilon(\vec{r},N,T) = 1+\chi(\vec{r}, \omega_0 ,N,T)
\end{equation}
is the relative permittivity, and
\begin{equation}
n_\mathrm{g}(\vec{r}) = n(\vec{r})+\omega_0 \partial_\omega n(\vec{r},\omega)|_{\omega = \omega_0}
\end{equation}
is the group index with $n=\sqrt{\varepsilon}$ being the refractive index.
Dispersion effects, not included in $n_\mathrm{g}$ (such as gain dispersion) could be modeled by the function  $\vec{D}_\mathrm{disp}(\vec{r},t)$ \cite{Wenzel2021}. In this paper, gain dispersion is neglected and $\vec{D}_\mathrm{disp}=0$ is assumed.

Spontaneous emission in the active layer is described by the Langevin force 
\begin{equation}
\vec{F}_\mathrm{sp}(\vec{r},t,N)=\frac{2}{\varepsilon_0c^2}\partial_t\vec{\mathcal{J}}_\mathrm{sp}(\vec{r},t,N)e^{-i\omega_0t}
\label{eq:Fsp}
\end{equation}
with the  stochastic current density $\vec{\mathcal{J}}_\mathrm{sp}$ that is real--valued, has zero mean and vanishes outside of the active layer.
Its spectral decomposition 
\begin{equation}
\vec{J}_\mathrm{sp}(\vec{r},\omega,N)=\int_{-\infty}^{+\infty}\vec{\mathcal{J}}_\mathrm{sp}(\vec{r},t,N)e^{-i\omega t}\,dt
\end{equation}
has the correlation function \cite{Wenzel1994,sondergaard2001general}
\begin{multline}
\langle J_\mathrm{sp,i}^*(\vec{r},\omega,N)J_\mathrm{sp,j}(\vec{r}',\omega',N) \rangle\\
=2\varepsilon_0\hbar\omega^2\Im\varepsilon|_\mathrm{cv}(\vec{r},\omega,N)n_\mathrm{sp}(\vec{r},\omega,N)\\
\times\delta_{ij}\delta(\vec{r}-\vec{r}')2\pi\delta(\omega-\omega')
\label{eq:correljj}
\end{multline}
following from the fluctuation--dissipation theorem \cite{marani1995spontaneous,Henry1996}.
Here, $\langle\cdot\rangle$ denotes the ensemble or temporal average, $i$ and $j$ are  the cartesian components of $\vec{J}_\mathrm{sp}$, $\varepsilon_0$ is the permittivity of free space, $\hbar$ is the reduced Planck's constant, $\Im\varepsilon|_\mathrm{cv}$ is the imaginary part of the relative permittivity due to transitions between the conduction and valence bands proportional to the optical gain in the active layer, and $n_\mathrm{sp}$ is the population inversion factor.

In order to determine $\langle F_{\mathrm{sp},i}^\ast(\vec{r},t)F_{\mathrm{sp,j}}(\vec{r}',t')\rangle$, the correlation function $\langle\partial_t\mathcal{J}_\mathrm{sp,i}(\vec{r},t)\partial_{t'}\mathcal{J}_\mathrm{sp,j}(\vec{r}',t')\rangle e^{-i\omega_0(t'-t)}$ has to be calculated. Using \eqref{eq:correljj} and considering that the integrand is rapidly oscillating except for $\omega=\omega_0$ as well as that $\omega^4\Im\varepsilon|_\mathrm{cv}(\vec{r},\omega)n_\mathrm{sp}(\vec{r},\omega)$ is slowly varying with $\omega$ compared to $\exp(i\omega(t'-t))$,
\begin{equation}
\begin{aligned}
&\langle\partial_t\mathcal{J}_\mathrm{sp,i}(\vec{r},t,N)\partial_{t'}\mathcal{J}_\mathrm{sp,j}(\vec{r}',t',N)\rangle e^{-i\omega_0(t'-t)}\\
&=\frac{1}{4\pi^2}\iint\omega\omega'\langle J_\mathrm{sp,i}^*(\vec{r},\omega,N)J_\mathrm{sp,i}(\vec{r}',\omega',N) \rangle\\
&\hspace{3.7cm}\times e^{-i(\omega-\omega_0)t+i(\omega'-\omega_0)t'}\,d\omega d\omega'\\
&=\frac{\varepsilon_0\hbar}{\pi}\int\omega^4\Im\varepsilon|_\mathrm{cv}(\vec{r},\omega,N)n_\mathrm{sp}(\vec{r},\omega,N)e^{i(\omega-\omega_0)(t'-t)}\,d\omega\\
&\hspace{6cm}\times\delta_{ij}\delta(\vec{r}-\vec{r}')\\
&\approx 2\varepsilon_0\hbar\omega_0^4\Im\varepsilon|_\mathrm{cv}(\vec{r},\omega_0,N)n_\mathrm{sp}(\vec{r},\omega_0,N)\delta_{ij}\delta(\vec{r}-\vec{r}')\delta(t-t')
\end{aligned}
\end{equation}
is gained.
The imaginary part of the permittivity which enters \eqref{eq:correljj} and vanishes outside the active layer can be written as
\begin{equation}
\label{eq:Imepscv}
\Im\varepsilon|_\mathrm{cv}(\vec{r},\omega_0,N)=\frac{n_\mathrm{a}(\vec{r},N)g_\mathrm{a}(\vec{r},N)}{k_0}
\end{equation}
where $g_\mathrm{a}$ is the local gain in the active layer and $n_\mathrm{a}$ the corresponding refractive index.
Due to the fact that $n_\mathrm{sp}$ approaches $\pm\infty$ at transparency where $g_\mathrm{a}$ vanishes as discussed in \cite{Wenzel2021}, it is advantageous to introduce the spontaneous emission per unit length
\begin{equation}
\label{eq:rsp}
r_\mathrm{sp}(\vec{r},N)=g_\mathrm{a}(\vec{r},N)n_\mathrm{sp}(\vec{r},\omega_0,N)
\end{equation}
which behaves smooth and can be approximated by an analytic function, see  \eqref{eq:rspN}.
Finally, the correlation function 
\begin{multline}
\langle F_{\mathrm{sp},i}^\ast(\vec{r},t,N)F_{\mathrm{sp,j}}(\vec{r}',t',N) \rangle\\
 \approx\frac{8\hbar\omega_0^3}{\varepsilon_0c^3}n_\mathrm{a}(\vec{r})r_\mathrm{sp}(\vec{r},N)\delta_{ij}\delta(\vec{r}-\vec{r}')\delta(t-t')
\label{eq:correlFF}
\end{multline}
is obtained in basic agreement with Ref. \cite{Shams2007}.

The complex--valued relative permittivity is written as
\begin{equation}
\varepsilon(\vec{r},N,T)=\varepsilon_\mathrm{s}(\vec{r})+\delta\varepsilon_\mathrm{s}(\vec{r})+\delta\varepsilon(\vec{r},N,T)
\label{eq:indexmodel}
\end{equation}
consisting of a real--valued part $\varepsilon_\mathrm{s}$ given by the structure defined by the compositions of the materials involved, a small complex--valued correction $\delta\varepsilon_\mathrm{s}$ due to absorption caused by doping, and the impact of $N$ and $T$ on $\varepsilon(\vec{r})$.
In this paper, isothermal conditions are considered. The dependence of $\delta\varepsilon$ on $N$ includes carrier induced change of the refractive index, free--carrier absorption and gain (\emph{i.e.} $\Im\varepsilon|_\mathrm{cv}$). In the following, the dependencies on the independent variables are not explicitly noted unless absolutely necessary.

\section{Coupled-wave model}

In a PCSEL, $\varepsilon_\mathrm{s}$(\vec{r}) is periodic in $x$ and $y$ within a finite domain $(x,y)\in[0,L]\times [0,L]$.
For a square lattice with lattice constant $a$ it can be expanded  into the Fourier series
\begin{equation}
\varepsilon_\mathrm{s}(\vec{r})=\overline{\varepsilon}_\mathrm{s}(z)+\sum_{m,n\ne 0}\xi_{m,n}(z)e^{-im\beta_0x-in\beta_0y}
\label{eq:expansion1}
\end{equation}
where
\begin{equation}
\beta_0=\frac{2\pi}{a}
\end{equation}
is the Bragg wave number,
\begin{equation}
\xi_{m,n}(z)=\frac{1}{a^{2}}\iint \varepsilon_\mathrm{s}(\vec{r})e^{im\beta_{0}x +in\beta_{0}y}dxdy 
\label{eq:xi}
\end{equation}
are the Fourier coefficients with $m$ and $n$ being the integer Fourier indices, and
\begin{equation}
\overline{\varepsilon}_\mathrm{s}(z)=\xi_{0,0}(z)=\frac{1}{a^{2}}\iint \varepsilon_\mathrm{s}(\vec{r})dxdy
\label{eq:nav}
\end{equation}
is the permittivity averaged in--plane.
The correction to the permittivity in \eqref{eq:indexmodel} is approximated as
\begin{equation}
\delta\varepsilon_\mathrm{s}(\vec{r})\approx\overline{\delta\varepsilon}_\mathrm{s}(z)=\frac{1}{a^{2}}\iint\delta \varepsilon_\mathrm{s}(\vec{r})dxdy
\label{eq:delta_n}
\end{equation}
neglecting the Fourier terms $m,n\ne 0$. 
Similarly, $\delta\varepsilon\approx\overline{\delta\varepsilon}$ and $n_\mathrm{g}\approx\overline{n}_\mathrm{g}$ holds.
In \eqref{eq:xi} -- \eqref{eq:delta_n}, the integration extends over one unit cell $(x,y)\in[-a/2,a/2]\times [-a/2,a/2]$ of the PC.

Typically, the PC layer is composed of two materials, namely air or another material with relatively low refractive index within periodically in both lateral directions repeating features (region I) and the surrounding semiconductor material (region II).
For perpendicular side walls of the features forming the PC
\begin{equation}
\label{eq:Axy}
\varepsilon(\vec{r})=\varepsilon_\mathrm{II}(z)+\Delta\varepsilon(z)A(x,y)
\end{equation}
holds (similarly for each of the two terms in \eqref{eq:indexmodel}) where 
\begin{equation}
\label{eq:A}
A(x,y)=
\begin{cases}
1 & \quad\text{for}\quad x,y\in \text{region I}\\
0 & \quad\text{for}\quad x,y\in \text{region II}
\end{cases}
\end{equation}
is the shape function dependent only on the in--plane coordinates $x$ and $y$.
The difference 
\begin{equation}
\label{eq:Deltan}
\Delta \varepsilon(z)=\varepsilon_\mathrm{I}(z) - \varepsilon_\mathrm{II}(z),
\end{equation}
with $\varepsilon_\mathrm{I}(z)$ and $\varepsilon_\mathrm{II}(z)$ being the permittivity in regions I and II, respectively,  is  non-vanishing only within a definite range.
If the surface after the regrowth of the PC layer containing the etched features is planar, $\Delta \varepsilon(z)=\text{constant}\ne0$ only within the PC layer $z\in [z_\mathrm{min},z_\mathrm{max}]$.
Otherwise,  $\Delta \varepsilon(z)\ne0$ also in the region $z>z_\mathrm{max}$ above the PC layer (seen from the growth perspective).

To achieve out--coupling along $z$, the periodicity has to enable second order Bragg diffraction.
Assuming TE--like polarization\footnote{In the PC community this type of polarization is called \lq TM\rq\ sometimes.}, the electric field strength \vec{E}(\vec{r},t) can then be  approximated by \cite{Liang2012}
\begin{equation}
\label{eq:Ansatz}
\begin{aligned}
E_{x}(\vec{r},t)=&\sqrt{\mathcal{N}}\Theta(z)\left[v^+(x,y,t)e^{-i\beta_0y}+v^-(x,y,t)e^{i\beta_0y}\right]\\
&+\Delta E_x(\vec{r},t)\\
&+\sum_{\sqrt{m^2+n^2}>1}E_{x,m,n}(\vec{r},t)e^{-im\beta_0x-in\beta_0y}\\
E_{y}(\vec{r},t)=&\sqrt{\mathcal{N}}\Theta(z)\left[u^+(x,y,t)e^{-i\beta_0x}+u^-(x,y,t)e^{i\beta_0x}\right]\\
&+\Delta E_y(\vec{r},t)\\
&+\sum_{\sqrt{m^2+n^2}>1}E_{y,m,n}(\vec{r},t)e^{-im\beta_0x-in\beta_0y}\\
E_{z}(\vec{r},t)=&0
\end{aligned}
\end{equation}
consisting of the basic waves due to second order diffraction (first term), vertically radiated waves $\Delta E_{x,y}$ due to first order diffraction (second term) and higher order waves (third term).
The first two terms are identical to those of second--order DFB lasers \cite{Shams2000}. 
The third term is unique to PCSELs.

The vertical mode $\Theta(z)$ solves the 1D eigenvalue problem
\begin{equation}
\label{eq:1Deigenvalue}
\frac{d^2\Theta}{dz^2}+k_0^2\overline{\varepsilon}_\mathrm{s}(z)\Theta(z)=\beta^2\Theta(z)
\end{equation}
subject to homogeneous Dirichlet boundary conditions at $z=0$ and $z=L_z$ corresponding to the lower boundary and upper boundary, respectively,  of the vertical waveguide and continuity for $\Theta(z)$ and its derivative at hetero boundaries appropriate for TE--like polarization. 
The real--valued propagation constant $\beta$ being the square root of the eigenvalue of \eqref{eq:1Deigenvalue} can be written as
\begin{equation}
\label{eq:neff}
\beta=n_\mathrm{eff}k_0
\end{equation}
defining the effective index $n_\mathrm{eff}$.
The mode profile $\Theta(z)$ is assumed to be normalized according to
\begin{equation}
\label{eq:norm}
\int_0^{L_z} |\Theta|^2\,dz=1.
\end{equation}
The normalization factor $\mathcal{N}$ in \eqref{eq:Ansatz} is chosen to be
\begin{equation}
\mathcal{N}=\frac{2k_0}{\varepsilon_0c\beta}=\frac{2}{\varepsilon_0cn_\mathrm{eff}}
\label{eq:normpower}
\end{equation}
so that $u^\pm$ and $v^\pm$ have the unit $\sqrt{\text{W}/\text{m}}$.
However, other choices are also possible.

Inserting \eqref{eq:Ansatz} into the \eqref{eq:Helmholtz}, making the usual slowly varying amplitude approximation (SWA)
including omission of all lateral derivatives of $E_{x|y,m,n}$, $|m|+|n|> 1$, neglecting rapidly varying terms, multiplying with $\Theta^\ast$, integrating along $z$,  and assuming that only the basic waves corresponding to  the first term in \eqref{eq:Ansatz}  are important in generating the radiative and higher order waves, the coupled--wave equations
\begin{multline}
\label{eq:CWT}
\frac{1}{v_\mathrm{g}}
\begin{pmatrix}
\partial_t u^+\\
\partial_t u^-\\
\partial_t v^+\\
\partial_t v^-
\end{pmatrix}
=\begin{pmatrix}
- \partial_x u^+\\
+ \partial_x u^-\\
- \partial_y v^+\\
+ \partial_y v^-
\end{pmatrix}
-\left[i\Delta\beta(x,y)-i\bm{\mathrm{C}}\right]\\
\times\begin{pmatrix}
u^+(x,y,t)\\
u^-(x,y,t)\\
v^+(x,y,t)\\
v^-(x,y,t)
\end{pmatrix}
+
\begin{pmatrix}
F_u^+(x,y,t)\\
F_u^-(x,y,t)\\
F_v^+(x,y,t)\\
F_v^-(x,y,t)
\end{pmatrix}
\end{multline}
can be derived \cite{Inoue2019}.
Here, $\partial_{x,y}$ denote the partial derivatives with respect to $x,y$.
The equations are to be solved on a square
$(x,y)\in[0,L]\times [0,L]$ subject to the boundary conditions
\begin{equation}
\label{eq:bc}
u^+(0,y,t)=u^-(L,y,t)=v^+(x,0,t)=v^-(x,L,t)=0.
\end{equation}
The complex--valued $4\times4$ field--coupling matrix $\bm{\mathrm{C}}$ is presented in Appendix \ref{sec:Cmatrix}.
The relative propagation factor reads
\begin{equation}
\label{eq:Deltabeta}
\Delta\beta=\frac{\beta^2-\beta_0^2}{2\beta_0}+\frac{k_0^2}{2\beta_0}\int_0^{L_z}
\left[\overline{\delta\varepsilon}_\mathrm{s}(z)+\overline{\delta\varepsilon}(z)\right]\vert \Theta\vert^{2}\,dz.
\end{equation}
In general, $\Delta\beta$ can depend on the in-plane coordinates $x$ and $y$ via
the function $\overline{\delta\epsilon}(z)$ in the second term of the  rhs of \eqref{eq:Deltabeta}, modeling spatially varying carrier density and temperature distributions due to 
non-uniform carrier injection, spatial holeburning, and heat flow, \emph{e.g.}, similarly as in \cite{zeghuzi2019}. 
In this paper, we assume a constant $\Delta\beta$ within the simulation domain.

In the what follows, the Bragg condition $\beta_0=\beta\equiv k_0n_\mathrm{eff}$ is employed.
The factor in front of the time--derivative is the inverse modal group velocity given by \cite{Snyder} 
\begin{equation}
\label{eq:vg}
\frac{1}{v_\mathrm{g}}=\frac{1}{cn_\mathrm{eff}}\int_0^{L_z} \overline{n}_\mathrm{s}(z)\overline{n}_\mathrm{g}(z)\vert \Theta(z)\vert^{2}\,dz,
\end{equation}
assuming $n\approx \overline{n}_\mathrm{s}=\sqrt{\overline{\varepsilon}_\mathrm{s}}$ in the integrand.
The new Langevin forces
\begin{equation}
\begin{pmatrix}
F_u^+\\
F_u^-\\
F_v^+\\
F_v^-
\end{pmatrix}
=\frac{i}{2\beta_0\sqrt{\mathcal{N}}}
\int_{\substack{\mathrm{active} \\ \mathrm{layer}}} F_\mathrm{sp}(\vec{r},t)\Theta^\ast(z)\,dz
\begin{pmatrix}
e^{+i\beta_0x}\\
e^{-i\beta_0x}\\
e^{+i\beta_0y}\\
e^{-i\beta_0y}
\end{pmatrix}
\label{eq:Fi}
\end{equation}
have the correlation functions
\begin{multline}
\label{eq:correlF}
\langle F_i^\ast(x,y,t)F_j(x',y',t')\rangle\\
=2D_{F^\ast F}\delta_{ij}\delta(x-x')\delta(y-y')\delta(t-t')
\end{multline}
with $i$ and $j$ denoting here the different components of \eqref{eq:Fi}.
The diffusion coefficient 
\begin{equation}
\label{eq:Dff1}
2D_{F^\ast F}
=\hbar\omega_0\frac{n_\mathrm{a}}{n_\mathrm{eff}}\Gamma r_\mathrm{sp}
\end{equation}
follows from \eqref{eq:correlFF}, the normalization \eqref{eq:normpower}, and the confinement factor 
\begin{equation}
\label{eq:Gamma}
\Gamma=\int_{\substack{\mathrm{active} \\ \mathrm{layer}}}\vert \Theta(z)\vert^{2}\,dz.
\end{equation}

The power emitted at a surface of the PCSEL at $z=z_\mathrm{out}$ where $z_\mathrm{out}=0-\delta$ or $z_\mathrm{out}=L_z+\delta$ ($\delta$ small positive number) with refractive index $n(x,y,z_\mathrm{out})$ can be calculated from the radiating waves as
\begin{multline}
P_\mathrm{out}(z_\mathrm{out},t)
=\frac{\varepsilon_0c}{2}\int_0^{L}\int_0^{L} n(x,y,z_\mathrm{out}) \\
\times\Big[|\Delta E_x(x,y,z_\mathrm{out},t)|^2
+|\Delta E_y(x,y,z_\mathrm{out},t)|^2\Big] \,dxdy.
\label{eq:Pout}
\end{multline}
According to \cite{Liang2013}, the radiating waves are the solutions of
\begin{equation}
\label{eq:radwaves0}
\begin{aligned}
& \frac{\partial^2 \Delta E_x}{\partial z^2} + k_0^2\left[\overline{\varepsilon}_\mathrm{s}(z)+\overline{\delta\varepsilon}_\mathrm{s}(z)\right] \Delta E_x(\vec{r},t)\\
& = -k_0^2\sqrt{\mathcal{N}}\left[\xi_{0,-1}(z)v^+(x,y,t)+\xi_{0,1}(z)v^-(x,y,t)\right]\Theta(z)\\
&\frac{\partial^2 \Delta E_y}{\partial z^2} + k_0^2\left[\overline{\varepsilon}_\mathrm{s}(z)+\overline{\delta\varepsilon}_\mathrm{s}(z)\right] \Delta E_y(\vec{r},t)\\
& = -k_0^2\sqrt{\mathcal{N}}\left[\xi_{-1,0}(z)u^+(x,y,t)+\xi_{1,0}(z)u^-(x,y,t)\right]\Theta(z)
\end{aligned}
\end{equation}
similarly as in a second order DFB laser \cite{Shams2000}.
These inhomogeneous equations can be solved by the Green's function approach,
\begin{equation}
\label{eq:radwaves1}
\begin{aligned}
\Delta E_x(\vec{r},t)=&k_0^2\sqrt{\mathcal{N}}v^+\int_{z_\mathrm{min}}^{z_\mathrm{max}}\xi_{0,-1}(z')G(z,z')\Theta(z')\,dz'\\
                  &+k_0^2\sqrt{\mathcal{N}}v^-\int_{z_\mathrm{min}}^{z_\mathrm{max}}\xi_{0,1}(z')G(z,z')\Theta(z')\,dz'\\
\Delta E_y(\vec{r},t)=&k_0^2\sqrt{\mathcal{N}}u^+\int_{z_\mathrm{min}}^{z_\mathrm{max}}\xi_{-1,0}(z')G(z,z')\Theta(z')\,dz'\\
                  &+k_0^2\sqrt{\mathcal{N}}u^-\int_{z_\mathrm{min}}^{z_\mathrm{max}}\xi_{1,0}(z')G(z,z')\Theta(z')\,dz'
\end{aligned}
\end{equation}
where the Green's function $G(z,z')$ is the solution of 
\begin{equation}
\label{eq:Greeneq}
\frac{d^2G(z,z')}{dz^2}+k_0^2[\overline{\varepsilon}_\mathrm{s}(z)+\overline{\delta\varepsilon}_\mathrm{s}(z)]G(z,z')=-\delta(z-z')
\end{equation}
subject to homogeneous Robin boundary conditions.
The near and far field distributions of the emitted beam are given in Appendix \ref{sec:NF_FF}.

\section{Spectral problem}

The spectral problem to calculate the laser modes is obtained from \eqref{eq:CWT} by setting $F_i=0$ ($i=1,2,3,4$) and inserting the Ansatz
\begin{equation}
\label{eq:Ansatz_spectral}
\begin{pmatrix}
u^+(x,y,t)\\
u^-(x,y,t)\\
v^+(x,y,t)\\
v^-(x,y,t)
\end{pmatrix}
=e^{i\Omega t}
\begin{pmatrix}
\Phi_{u}^+(x,y)\\
\Phi_{u}^-(x,y)\\
\Phi_{v}^+(x,y)\\
\Phi_{v}^-(x,y)
\end{pmatrix}
\end{equation}
where $\Omega$ is the complex--valued relative frequency serving as eigenvalue. 
Let the solutions (modes) of the spectral problem (see Appendix \ref{sec:adjoint_ortho}, equation \eqref{eq:spec}) be distinguished by the index $\mu$ and let  the mode $\mu=1$ be the lasing mode for which  $\text{Im}(\Omega_{\mu=1})=0$ holds. 
This threshold  condition can be fulfilled by varying 
some parameter, such as the modal gain 

\begin{equation}
g=\frac{n_\mathrm{a}}{n_\mathrm{eff}}\Gamma g_\mathrm{a},
\label{eq:gmod1}
\end{equation}
yielding the threshold gain $g_\mathrm{th}$. This relation follows from \eqref{eq:Imepscv} and \eqref{eq:Deltabeta}. 
Another parameter of interest is the additional gain 
\begin{equation}
\Delta g_\mu = \frac{2}{v_\mathrm{g}}\Im(\Omega_{\mu\ne 1})
\label{eq:Deltag}
\end{equation}
needed by the other modes to reach threshold.
The relative wavelengths of the modes including the  threshold wavelength are  given by
\begin{equation}
\label{eq:lambdath}
\Delta\lambda_\mu=\frac{d\lambda}{d\omega}\bigg|_{\lambda_0}\Re(\Omega_\mu).
\end{equation}
For spatially uniform $\Delta\beta=\overline{\Delta\beta}$ as assumed here a new eigenvalue
\begin{equation}
\tilde{\beta}_\mu=\left(\overline{\Delta\beta}+\frac{\Omega_\mu}{v_\mathrm{g}}\right)
\end{equation}
can be introduced. 
The imaginary part of the eigenvalue $\tilde{\beta}$ delivers the out--coupling loss (due to out--coupled radiation at the edges and towards the surfaces).
The threshold gain is then obtained from
\begin{equation}
\label{eq:gth}
g_\mathrm{th}=2\Im\tilde{\beta}_{\mu=1}+\alpha
\end{equation}
where $\alpha$ summarizes all internal modal losses.

Besides the threshold gain from which the threshold current $I_\mathrm{th}$ can be determined, the laser characteristics involves also the slope efficiency \cite{Wenzel2017}
\begin{equation}
\label{eq:S}
S=\frac{P_\mathrm{out}}{I-I_\mathrm{th}}=\frac{\hbar\omega}{q}\eta_\mathrm{i}\eta_\mathrm{d}.
\end{equation}
Here, $I$ is the injection current, $\omega$ is the lasing (angular) frequency, $q$ is the elementary charge, 
\begin{equation}
\label{eq:etai}
\eta_\mathrm{i}=\frac{q}{\hbar\omega}\frac{P_\mathrm{st}}{I-I_\mathrm{th}}
\end{equation}
is the internal efficiency to be determined above threshold, and 
\begin{equation}
\label{eq:etad}
\eta_\mathrm{d}=\frac{P_\mathrm{out}}{P_\mathrm{st}}
\end{equation}
is the  external differential efficiency which can be calculated at threshold, \emph{c.f} \cite{noll1990analysis}.
The power generated by stimulated emission within the cavity entering \eqref{eq:etad} is given by
\begin{equation}
\label{eq:Pstim}
P_\mathrm{st}=\int_{\substack{\mathrm{active} \\ \mathrm{layer}}}\int_0^L\int_0^L Q_\mathrm{st}(\vec{r})\,dxdydz
\end{equation}
with the power density \cite{Landau} 
\begin{equation}
\label{eq:Qstim}
Q_\mathrm{st}=\frac{\varepsilon\omega_0}{2}
\Im \varepsilon|_\mathrm{cv}\vec{E}\vec{E}^\ast
\end{equation}
where $\vec{E}(\vec{r},t)$ is the complex--valued slowly temporally--varying field strength 
defined in \eqref{eq:Ansatz_time}.

In what follows, the single--mode approximation 
\begin{equation}
\label{eq:single_mode}
\begin{pmatrix}
u^+(x,y,t)\\
u^-(x,y,t)\\
v^+(x,y,t)\\
v^-(x,y,t)
\end{pmatrix}
=
f(t)\begin{pmatrix}
\Phi_u^+(x,y)\\
\Phi_u^-(x,y)\\
\Phi_v^+(x,y)\\
\Phi_v^-(x,y)
\end{pmatrix}
\end{equation}
is employed where $f(t)$ is the mode amplitude and $\Phi_{u|v}^\pm$ is a solution of the spectral problem \eqref{eq:spec}.
Inserting \eqref{eq:Imepscv} and the first term of \eqref{eq:Ansatz} into \eqref{eq:Qstim}, considering \eqref{eq:single_mode} and the normalization \eqref{eq:normpower} 
 yields
\begin{equation}
\label{eq:Qstim1}
Q_\mathrm{st}=
\frac{n_\mathrm{a}}{n_\mathrm{eff}}g_\mathrm{a}|\Theta|^2\langle |f|^2\rangle\vert \vec{\Phi} \vert^2
\end{equation}
where rapidly varying terms were neglected,
$\langle |f|^2 \rangle$ denotes the mean intensity of the lasing mode,
and
\begin{equation}
\vert \vec{\Phi} \vert^2=|\Phi_u^+|^2+|\Phi_v^+|^2+|\Phi_u^-|^2+|\Phi_v^-|^2.
\end{equation}
Inserting \eqref{eq:Qstim1} into \eqref{eq:Pstim}, employing \eqref{eq:gmod1} , and the definition of the confinement factor \eqref{eq:Gamma} yields
\begin{equation}
\label{eq:PstimPCSELN}
P_\mathrm{st} = \langle |f|^2\rangle \iint g\vert \vec{\Phi} \vert^2\,dxdy
\end{equation}
where the integration extends over $(x,y)\in[0,L]\times [0,L]$.
Finally, the external differential efficiency is obtained by inserting 
\eqref{eq:PstimPCSELN} and \eqref{eq:Pout}  into \eqref{eq:etad}  
accounting for \eqref{eq:single_mode} which allows eliminating $\langle |f|^2\rangle$. 
The modal gain has to be taken at threshold.

\section{Spectral linewidth}

The spectral linewidth of a semiconductor laser is given by  \cite{henry1986phase}, \cite{Tromborg1991}
\begin{equation}
\label{eq:nu2}
\Delta\nu=\frac{KR_\mathrm{sp}}{4\pi I_\mathrm{ph}}\left(1+\alpha_\mathrm{H,eff}^2\right).
\end{equation}
Using the inner product \eqref{eq:inner} derived in the Appendix \ref{sec:adjoint_ortho} and making the single mode approximation \eqref{eq:single_mode}, the results of Ref. \cite{Wenzel2021} can be adopted to a PCSEL to calculate the different quantities entering \eqref{eq:nu2}.
After assuming constant $v_{\mathrm{g}}$ and neglecting gain dispersion and compression, we can write the  intra--cavity photon number, the rate of spontaneous emission into the lasing mode, and the
longitudinal excess factor of spontaneous emission (Petermann factor) as 
\begin{equation}
\label{eq:photonnumber}
I_\mathrm{ph}=\frac{\langle |f|^2 \rangle}{\hbar\omega v_\mathrm{g}}\iint\vert \vec{\Phi} \vert^2\,dxdy,
\end{equation}
\begin{equation}
\begin{aligned}
\label{eq:Rsp}
R_\mathrm{sp}&=\frac{2v_\mathrm{g}}{\hbar\omega}\frac{\iint\vert \vec{\Phi} \vert^2D_{F^\ast F}\,dxdy}{\iint\vert \vec{\Phi} \vert^2\,dxdy}\\
&=\frac{v_\mathrm{g}n_\mathrm{a}}{n_\mathrm{eff}}\frac{\iint\vert \vec{\Phi} \vert^2 \Gamma r_\mathrm{sp} \,dxdy}{\iint\vert \vec{\Phi} \vert^2\,dxdy},
\end{aligned}
\end{equation}
and
\begin{equation}
\label{eq:Petermann}
K=\frac{\left(\iint\vert \vec{\Phi} \vert^2\,dxdy\right)^2}{\left|\left(\vec{\Phi},\vec{\Phi}\right)\right|^2}
\end{equation}
with $\left(\vec{\Phi},\vec{\Phi}\right)$ given in \eqref{eq:inner},
respectively.
The rate of stimulated recombination needed to calculate the effective linewidth enhancement (Henry) factor $\alpha_\mathrm{H,eff}$ by the expression given in \cite{Wenzel2021} (equation (78) there) is obtained
from the power density \eqref{eq:Qstim1}
divided by the energy $\hbar\omega=\hbar[\omega_0+\text{Re}(\Omega]$ exchanged per recombination event,
\begin{equation}
R_\mathrm{st}=\frac{Q_\mathrm{st}}{\hbar\omega}.
\end{equation}
In order to use the expression for $\alpha_\mathrm{H,eff}$  given in \cite{Wenzel2021}, $R_\mathrm{st}$ has to be averaged
over the active layer with thickness $d$,
\begin{equation}
\overline{R}_\mathrm{st}=\frac{1}{d}\int_{\substack{\mathrm{active} \\ \mathrm{layer}}} R_\mathrm{st}(z)\,dz=\frac{g\langle |f|^2\rangle\vert \vec{\Phi} \vert^2}{d\hbar\omega}
\end{equation}
with the modal gain \eqref{eq:gmod1},
so that the derivative
\begin{equation}
\frac{\partial \overline{R}_\mathrm{st}}{\partial \langle |f|^2\rangle}
=\frac{g\vert \vec{\Phi} \vert^2}{d\hbar\omega}
\end{equation}
is obtained.
Assuming a constant differential carrier lifetime, it follows that
\begin{equation}
\label{eq:alphaHeff}
\alpha_\mathrm{H,eff}
= -\frac{\Re{(h_\alpha)}}{\Im{(h_\alpha)}}
\end{equation}
with
\begin{equation}
\label{eq:halpha}
h_\alpha=\frac{\left(\vec{\Phi},\frac{\partial \Delta\beta}{\partial N}g\vert \vec{\Phi}\vert^2\vec{\Phi}\right)}{\left(\vec{\Phi},\vec{\Phi}\right)}
\end{equation}
holds.
Equation \eqref{eq:nu2} has to be evaluated above threshold together with the governing equations for the averaged carrier density in order to determine $\langle |f|^2 \rangle$ as done in 
\cite{Wenzel2021} by solving the time--dependent coupled wave equations for a DBR laser.

At threshold, a useful quantity that can be calculated is the slope of the dependence of the linewidth on the inverse output power, i.e. the linewidth--power product 
\begin{equation}
\label{eq:linewidthPower}
\Delta\nu P_\mathrm{out}
=\frac{\eta_\mathrm{d}\hbar\omega v_\mathrm{g}KR_\mathrm{sp}G}{4\pi}
\left(1+\alpha_\mathrm{H,eff}^2\right)
\end{equation}
which is independent of $\langle |f|^2 \rangle$ if spatial holeburing effects are ignored.
Here, 
\begin{equation}
\label{eq:}
G=\frac{\iint g\vert \vec{\Phi} \vert^2\,dxdy}{\iint\vert \vec{\Phi} \vert^2\,dxdy}
=\frac{n_\mathrm{a}}{n_\mathrm{eff}}\frac{\iint \Gamma g_\mathrm{a}\vert \vec{\Phi} \vert^2\,dxdy}{\iint\vert \vec{\Phi} \vert^2\,dxdy}
\end{equation}
is the in--plane averaged gain and \eqref{eq:etad} and \eqref{eq:PstimPCSELN} was used. 

\begin{table}[!t]
\caption{Index ($\#$), thickness ($d$) and permittivity in regions I and II of the different layers \label{tab:layers}}
\centering
$\begin{array}{|d{2.0}|d{1.3}|d{2.3}|d{2.9}|d{2.3}|d{2.9}|}
\hline
  \multicolumn{1}{|c|}{{\#}} & \multicolumn{1}{c|}{{d}}   & \multicolumn{1}{c|}{{\varepsilon_\mathrm{s,I}}} & \multicolumn{1}{c|}{{\Im\delta\epsilon_\mathrm{s,I}}} & \multicolumn{1}{c|}{{\varepsilon_\mathrm{s,II}}} & \multicolumn{1}{c|}{{\Im\delta\epsilon_\mathrm{s,II}}}\\
  &\multicolumn{1}{c|}{{(\mu\text{m})}}  & &  & & \\
\hline\hline
1 &  0.1 & 12.065 & -1.066\cdot 10^{-3} & 12.065 & -1.066\cdot 10^{-3} \\
2 &  0.5 & 10.184 & -4.348\cdot 10^{-4} & 10.184 & -4.348\cdot 10^{-4} \\
3 &  1.0 & 10.186 & -2.174\cdot 10^{-4} & 10.186 & -2.174\cdot 10^{-4} \\
4 &  1.0 & 10.186 & -1.087\cdot 10^{-4} & 10.186 & -1.087\cdot 10^{-4} \\
5 &  0.3 & 10.187 & -4.348\cdot 10^{-5} & 10.187 & -4.348\cdot 10^{-5} \\
6 &  0.2 & 10.187 & -2.174\cdot 10^{-5} & 10.187 & -2.174\cdot 10^{-5} \\
7 & 0.075 & 12.096 & -2.369\cdot 10^{-6} & 12.096 & -2.369\cdot 10^{-6} \\
8 & 0.005 & 13.490 & -2.502\cdot 10^{-6} & 13.490 & -2.502\cdot 10^{-6} \\
9 &  0.05 & 12.096 & -7.111\cdot 10^{-6} & 12.096 & -7.111\cdot 10^{-6} \\
10&  0.1 & 12.096 & -7.107\cdot 10^{-5} & 12.096 & -7.107\cdot 10^{-5} \\\hline
\multicolumn{6}{|c|}{\text{air--hole}}\\\hline
11& 0.15  & 1.0 & 0.0 & 12.096 & -7.107\cdot 10^{-5} \\\hline
\multicolumn{6}{|c|}{\text{InGaP}}\\\hline
11& 0.15  & 10.115 & -6.500\cdot 10^{-4} & 12.096 & -7.107\cdot 10^{-5} \\\hline
12& 0.1 & 12.096 & -7.107\cdot 10^{-5} & 12.096 & -7.107\cdot 10^{-5} \\
13&	0.1 & 10.187 & -6.522\cdot 10^{-5} & 10.187 & -6.522\cdot 10^{-5} \\
14&  0.1 & 10.187 & -1.304\cdot 10^{-4} & 10.187 & -1.304\cdot 10^{-4} \\
15&  0.1 & 10.186 & -3.261\cdot 10^{-4} & 10.186 & -3.261\cdot 10^{-4} \\
16&  0.1 & 10.185 & -6.522\cdot 10^{-4} & 10.185 & -6.522\cdot 10^{-4} \\
17&  0.7 & 10.183 & -1.304\cdot 10^{-3} & 10.183 & -1.304\cdot 10^{-3} \\
18&  0.1 & 12.053 & -1.421\cdot 10^{-2} & 12.053 & -1.421\cdot 10^{-2} \\
\hline
\end{array}$
\end{table}

For a computation of $\Delta\nu P_\mathrm{out}$ expressions for the dependence of $r_\mathrm{sp}$, $g_\mathrm{a}$ and $\partial \Delta\beta/\partial N$ on $N$ are needed.
For simplicity, analytic expressions are used. The relation between the carriers injected into the active layer (responsible for stimulated emission) and the carrier densities in the barriers and bulk regions as studied in \cite{boni2023} is not addressed here.
The spontaneous emission per length introduced in \eqref{eq:rsp} is given by \cite{Wenzel2021}
\begin{equation}
\label{eq:rspN}
r_\mathrm{sp}(N)=g^\prime_NN_\mathrm{tr}\ln\sqrt{1+\left(\frac{N}{N_\text{tr}}\right)^2}
\end{equation}
and the local gain in the active layer by
\begin{equation}
\label{eq:gN}
g_\mathrm{a}(N)= g^\prime_NN_\text{tr}\ln\left[\frac{\max(N,N_\mathrm{cl,g})}{N_\text{tr}}\right].
\end{equation}
Here, $g^\prime_N$ is the differential gain, $N_\text{tr}$ the transparency carrier density, and $N_\mathrm{cl,g}$ the gain clamping carrier density.
At limit cases, equation \eqref{eq:rspN} exhibits the correct behavior $\lim_{N \to 0}r_\mathrm{sp}(N)\propto N^2$ and $\lim_{N \to \infty}r_\mathrm{sp}(N) = g(N)$.
The relative propagation factor $\Delta\beta$ which is used when calculating $\partial\Delta\beta/\partial N$
and when determining the threshold carrier density $N_\mathrm{th}$ is modeled by
\begin{multline}
\label{eq:Deltabeta1}
\Delta\beta(N) =k_0 \Delta n_N(N)\\
+\frac{i}{2}\left[g(N)-\alpha_0-\alpha_{\mathrm{WG}}-\alpha_N(N)\right]
\end{multline}
with the carrier--induced index change
\begin{equation}
\Delta n_N=-\frac{n_\mathrm{a}}{n_\mathrm{eff}}
\Gamma\tilde{\alpha}_\mathrm{H} g_N^\prime\sqrt{\frac{\max(N,N_\mathrm{cl,i})}{N_\mathrm{tr}}},
\end{equation}
with the scattering loss $\alpha_0$,
the modal waveguide absorption loss
\begin{equation}
\label{eq:alphWG}
\alpha_\mathrm{WG}=-\frac{k_0}{n_\mathrm{eff}}\int_0^{L_z}\Im\overline{\delta\varepsilon}_\mathrm{s}(z) \vert \Theta(z)\vert^{2}dz,
\end{equation}
and the modal loss due to free--carrier absorption in the active layer
\begin{equation}
\label{eq:alphaN}
\alpha_N(N)=\frac{n_\mathrm{a}}{n_\mathrm{eff}}\Gamma f_N N.
\end{equation}
Here, $\tilde{\alpha}_\mathrm{H}$ is a prefactor, $N_\mathrm{cl,i}$ the index clamping carrier density,
and $f_N$ the cross--section of free--carrier absorption.
Note, that sometimes the ratio $n_\mathrm{a}/n_\mathrm{eff}$ is included in $\Gamma$.
Equations \eqref{eq:rspN}--\eqref{eq:gN} and $\partial\Delta\beta/\partial N$ have to be computed at the threshold carrier density
$N_\mathrm{th}$ which is the solution of 
\begin{equation}
\label{eq:Nth}
\Im\tilde{\beta}=\Im\Delta\beta(N_\mathrm{th}).
\end{equation}
Finally, the threshold current is
\begin{equation}
\label{eq:Ith}
I_\mathrm{th}=qdL^2\left(\frac{N_\mathrm{th}}{\tau_\text{SRH}}+BN_\mathrm{th}^2+CN_\mathrm{th}^3\right)
\end{equation}
where $\tau_\text{SRH}$ is the carrier lifetime due to Shockley--Read--Hall recombination and $B$ and $C$ are the coefficients for spontaneous radiative and Auger recombination, respectively.

\begin{table}[!t]
\caption{Global parameters\label{tab:global}}
\centering
$\begin{array}{|c|c|c|}
\hline
\text{parameter} & \text{value} & \text{unit} \\
\hline\hline
\lambda_0	& 1.07\cdot 10^{-6} &\text{m}\\  
\hline
\beta_0	& k_0n_\mathrm{eff} & \\  
\hline
n_\mathrm{g,eff}=c/v_\mathrm{g}	& 3.8 & \\  
\hline
d	& 5\cdot 10^{-9} & \text{m}\\  
\hline
n_\mathrm{a}	& 3.673 & \\  
\hline
\tilde{\alpha}_\mathrm{H}       & 1 &\\ 
\hline
g^\prime			&	1.7\cdot 10^{-19} &	\text{m}^2\\
\hline  
N_\text{tr}		& 1.7\cdot 10^{24} &	\text{m}^{-3}\\ 
\hline
N_\mathrm{cl,g}=N_\mathrm{cl,g}	& 10^{22} &	\text{m}^{-3}\\ 
\hline
\alpha_0	& 0	& \text{m}^{-1}\\
\hline
f_N	& 16\cdot 10^{-22}	& \text{m}^{2}\\
\hline
\tau_\text{SRH}	& 1.5\cdot 10^{-9} 		& \text{s} \\ 
\hline
  B				      & 1.6\cdot10^{-16}	  &\text{m}^3\text{s}^{-1} \\ 
\hline
  C				      & 4\cdot10^{-42}		  &\text{m}^6\text{s}^{-1} \\
\hline
\end{array}$
\end{table}

\section{Numerical results}

The unit cell of the PCSELs under investigation is schematically shown in Fig. \ref{fig:PCSEL_schematic}b). It comprises a right--angled isosceles triangle (RIT) with reflection symmetry along $x=y$ chosen here as an example as this simple unit cell has been used to realize Watt class PCSELs \cite{hirose2014watt}. 
The RITs consists of air--holes or InGaP and are surrounded by GaAs.
The length $l$ of the legs of the RITs in units of the lattice constant $a$ is varied.
The thickness and the permittivity of each layer is given in Tab. \ref{tab:layers} where layer $\#11$ contains the etched RITs .
The active layer is layer $\#8$.
Note that the p-DBR depicted in Fig. \ref{fig:PCSEL_schematic}a is not considered and that the layers $\#1$ and $\#18$ corresponding to the n--substrate and the p--contact layer, respectively,  are assumed to be infinitely extended to avoid interference effects caused by reflections from semiconductor/air or semiconductor/metal interfaces.
Regions I and II are defined in \eqref{eq:A}.

The global parameters which are identical for both PCSEL types can be found in Tab. \ref{tab:global}.
It could be suspected that the effective recombination lifetime $\tau_\text{SRH}$ of an air--hole PCSEL is reduced due to additional non--radiative recombination at the air--semiconductor interface.
The dependence of gain and refractive index on the carrier density was obtained from a microscopic model for the permittivity \cite{wenzel1999improved}.
Note, that by setting $\alpha_0=0$ any additional losses such as from etching and regrowth, or scattering losses (possibly crucial for air--hole PCSELs due to the high index contrast between air and semiconductor) are not considered.
The lengths of the sides of the square active region equal to the simulation domain are varied within the range $150~\mu\text{m} \le L \le 800~\mu\text{m}$. For smaller lengths $L$, the mode intensity $\vert \vec{\Phi} \vert^2$ of the InGaP--based PCSEL is concentrated at the
lateral borders of the domain resulting in high threshold gain and low external differential efficiency because most of the radiation escapes through the lateral borders \cite{Mindaugas2024}. 

The numerical solution of \eqref{eq:1Deigenvalue}, the computation of the coupling matrix $\bm{\mathrm{C}}$ and the numerical solution of the spectral problem \eqref{eq:spec} with a finite--difference method based on a 4th-order scheme is detailed in \cite{Mindaugas2024, Mindaugas2023}.
The sum in \eqref{eq:Ansatz} is restricted by $\sqrt{m^2+n^2}\le 200$.  
The simulation domain is discretized into $20\times 20$ equal cells. 

\begin{figure}[!t]
\centering
\includegraphics[width=3in]{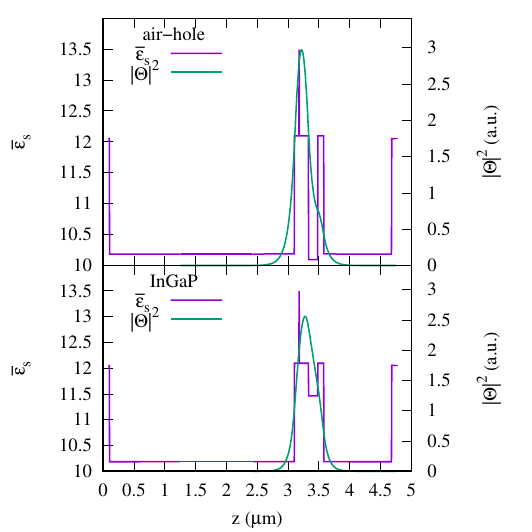}
\caption{Vertical profiles of the in-plane averaged structural permittivity (left axis) and mode intensity (right axis) of air--hole (top) and InGaP (bottom) based PCSELs. Lengths of legs are $0.6a$ and $0.8a$ for air--hole and InGaP, respectively, RITs.}
\label{fig:eps_mode}
\end{figure}

\begin{figure}[!t]
\centering
\includegraphics[width=3in]{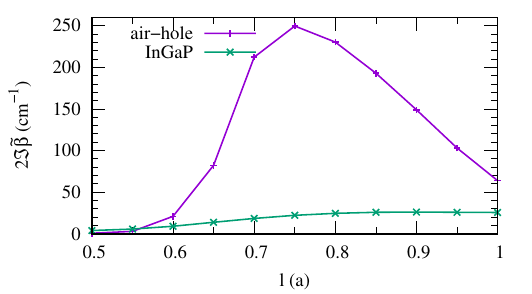}
\caption{Out--coupling loss of air--hole (violet) and InGaP (green) based PCSELS with $L=500~\mu$m versus the lengths of the legs of the RIT in unit of the lattice constant.}
\label{fig:a_simulation_data_loss}
\end{figure}

Fig.~\ref{fig:eps_mode} shows the vertical profiles of the in-plane averaged structural permittivity and of the mode intensity
of air--hole ($l=0.6a$) and InGaP ($l=0.8a$) based PCSELs. Further results of the solution of \eqref{eq:1Deigenvalue} are summarized in Tab. \ref{tab:results}. 
The InGaP--PCSEL is distinguished by a smaller overlap of the optical field with the active layer, due to the higher permittivity of InGaP compared to air which implies less steep gradient of $|\Theta|^2$ within the PC layer and, thus, a broader and less high peak of the vertical mode, and by larger modal waveguide absorption loss $\alpha_\mathrm{WG}$ due to the large value of $\Im\delta\epsilon_\mathrm{s,I}$ in layer $\#11$.

The elements of the coupling matrices $\bm{\mathrm{C}}$ for the air--hole PCSEL ($l=0.6a$) and for the InGaP PCSEL ($l=0.8a$) are collected in Tab. \ref{tab:coupling_matrix}.  
Due to the reduced permittivity contrast between InGaP and GaAs, the 2D coupling coefficients such as $C_{13}$ and $C_{14}$ of the InGaP PCSEL are by one order of magnitude smaller than those of the air--hole PCSEL. 
Contributions of the 1D coupling matrices $\mathbf{C}_{\mathrm{1D}}$ to total coupling matrices $\mathbf{C}$, however, remain of the same order.

\begin{table}[!t]
\caption{Results of the solution of the vertical waveguide equation and the spectral problem for $L=500~\mu$m \label{tab:results}}
\centering
$\begin{array}{|c|c|c|c|}
\hline
\text{parameter} & \text{air--hole} & \text{InGaP} & \text{unit}\\
l & 0.6 & 0.8 & a\\ 
\hline\hline
 n_\mathrm{eff} & 3.328 & 3.374 &\\  
 \Gamma & 0.014 &  0.0099 &\\  
 \alpha_\mathrm{WG} & 0.75 & 1.85 & \text{cm}^{-1} \\  
 \alpha_N & 0.62 & 0.64 & \text{cm}^{-1}\\  
 g_\mathrm{th} & 22.3 & 27.1 & \text{cm}^{-1}\\ 
 \Delta\lambda_\mathrm{th}=\lambda_1-\lambda_0 & -1.53 & 1.60 & \text{nm}\\ 
 N_\mathrm{th} & 2.8 & 4.1 & 10^{18}\text{cm}^{-3}\\ 
 I_\mathrm{th} & 0.65 & 1.12 & \text{A}\\
\theta_{x,\mathrm{FWHM}} & 0.15 & 0.15 & \text{degr.}\\
\theta_{y,\mathrm{FWHM}} & 0.15 & 0.15 & \text{degr.}\\
\Delta g_2=g_{2}-g_{\mathrm{th}} & 2.3 & 1.5 & \text{cm}^{-1}\\
\Delta\lambda_2=\lambda_2-\lambda_0 & -1.51 & 1.45 & \text{nm}\\
\hline
\end{array}$
\end{table}

\begin{table}[!t]
\caption{Coupling matrix \bm{\mathrm{C}} (cm$^{-1}$) \label{tab:coupling_matrix}}
\centering
$\begin{array}{|c|c|c|c|}
\hline
\multicolumn{4}{|c|}{\text{air--hole}\quad l=0.6a}\\
\hline
 776 + 206i & -604 + 768i &  245 -   0i &  286 -  47i \\
-197 - 704i &  776 + 206i &  286 +  47i &  245 +   0i \\
 245 +   0i &  286 -  47i &  774 + 207i & -605 + 773i \\
 286 +  47i &  245 -   0i & -197 - 711i &  774 + 207i \\
\hline\hline
\multicolumn{4}{|c|}{\text{InGaP}\quad l=0.8a}\\
\hline
  50 +  14i & -330 - 147i &  10  -   0i &   27 -   9i \\ 
-306 + 133i	&   50 +  14i &  27  +   9i &   10 +   0i \\
  10 +   0i	&   27 -   9i &  50  +  14i & -330 - 147i \\
  27 +   9i &   10 -   0i & -306 + 132i &   50 +  14i \\
\hline
\end{array}$
\end{table}

The optimal values of the lengths of the legs of the RIT were determined by calculating the dependence of the out--coupling loss $2\Im\tilde{\beta}$ on the lengths $l$ of the RIT legs, the results of which are shown in Fig. \ref{fig:a_simulation_data_loss} for $L=500~\mu$m.
The out--coupling loss of the air--hole PCSEL reaches very high values $2\Im\tilde{\beta}$$>200~\text{cm}^{-1}$ with a maximum around $l=0.75a$. 
In contrast, $2\Im\tilde{\beta}$ of the InGaP PCSEL rises only moderately with increasing values of $l$.

In what follows, $l=0.6a$ for the air--hole PCSEL and $l=0.8a$ for the InGaP PCSEL is chosen.
Further results for these lengths of the legs and $L=500~\mu$m are summarized in Tab. \ref{tab:results}.
Threshold carrier density $N_\mathrm{th}$ and current $I_\mathrm{th}$ of the air--hole PCSEL are lower because of the lower threshold gain $g_\mathrm{th}$ and the higher confinement factor $\Gamma$. The far--field divergence angles $\theta_{x,\mathrm{FWHM}}$ and $\theta_{y,\mathrm{FWHM}}$ (full width at half maximum, FWHM) for radiating into air ($n_\mathrm{out}=1$) are nearly identical for both PCSEL types and for both directions, the latter indicating a circular far--field distribution.
The lower threshold gain difference $\Delta g_1$ for the InGaP PCSEL can be improved through unit cell design, \emph{e.g.} with a double lattice as in \cite{Inoue2022}, which is, however, not in the scope of this paper.

\begin{figure}[!t]
\centering
\includegraphics[width=3in]{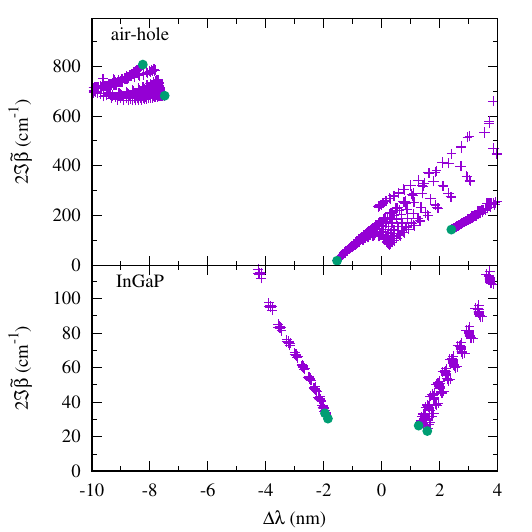}
\caption{Out--coupling loss versus the relative wavelength of the modes of air--hole (top) and InGaP (bottom) based PCSELs for $L=500~\mu$m. The losses and relative wavelengths originating from the eigenvalues of the coupling matrix are denoted by green bullets. Lengths of legs are $0.6a$ and $0.8a$ for air--hole and InGaP RITs, respectively.}
\label{fig:loss_lam}
\end{figure}

\begin{figure}[!t]
\centering
\includegraphics[scale=0.8]{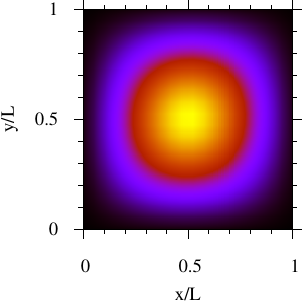}
\includegraphics[scale=0.8]{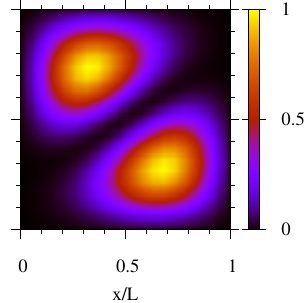}
\caption{Normalized intensity $\vert \vec{\Phi} \vert^2$ of the fundamental (left) and first higher order (right) mode of the air--hole PCSEL for $l=0.6a$ and $L=500~\mu$m.}
\label{fig:Air_GaAs_planar_mode_function}
\end{figure}

\begin{figure}[!t]
\centering
\includegraphics[scale=0.8]{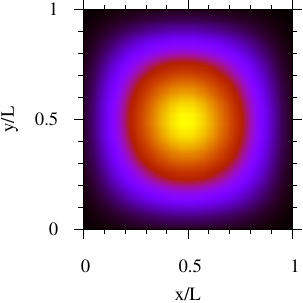}
\includegraphics[scale=0.8]{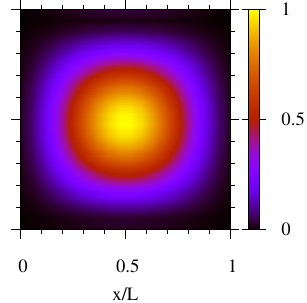}
\caption{Normalized intensity $\vert \vec{\Phi} \vert^2$  of the fundamental (left) and first higher order (right) mode of the InGaP PCSEL for $l=0.8a$ and $L=500~\mu$m.}
\label{fig:InGaP_GaAs_planar_mode_function}
\end{figure}

In Fig. \ref{fig:loss_lam} the complex--valued eigenvalues of the spectral problem and the coupling matrix are plotted in terms of out--coupling loss versus relative wavelength. 
In Refs. \cite{Liang2011,hirose2014watt} the 4 photonic bands in the vicinity of the second order $\Gamma$ point of the Brillouin zone of the PC  (corresponding to the 4 eigenvalues of the coupling matrix) are referred to as A--D with increasing frequency (decreasing wavelength).
It can be seen that the lasing modes, i.e. the modes with the lowest losses, correspond to photonic bands B and A for the air--hole PCSEL and the InGaP PCSEL, respectively.
This holds for the chosen lengths of the legs within the complete range $150 \le L/\mu$m $\le 800$ investigated.
The intensity distributions, normalized to maximum equals one, of the fundamental modes of both PCSEL types shown in Figs. \ref{fig:Air_GaAs_planar_mode_function} and \ref{fig:InGaP_GaAs_planar_mode_function} are nearly circular with a central peak. 
The first higher order mode of the air--hole PCSEL correspond to the photonic band B, too, and therefore its intensity distribution is different from that of the fundamental mode distinguished by two peaks and a zero between them. 
In contrast, the first higher order mode of the InGaP PCSEL belongs to another photonic band than the fundamental mode, namely B 
(\emph{c.f.} the relative wavelength $\Delta\lambda_2$ for the first higher order mode in Tab. \ref{tab:results}).
Therefore, the first higher order mode has also a circular intensity distribution albeit the field is polarized differently (not shown here).

\begin{figure}[!t]
\centering
\includegraphics[width=3in]{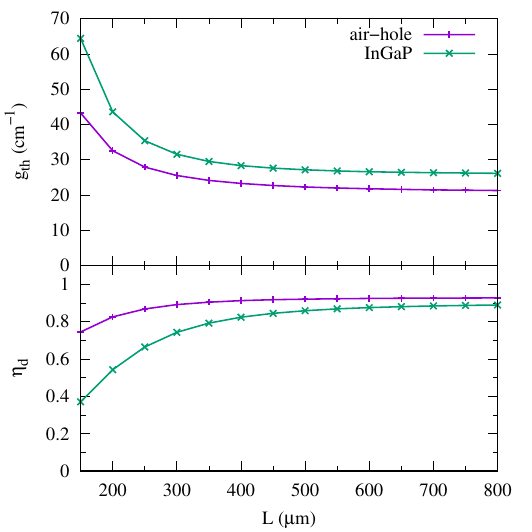}
\caption{Threshold gain (top) and total external differential efficiency (bottom) of air--hole (violet) and InGaP (green) based PCSELS versus the lengths of the sides of the active area. Lengths of RIT legs  are $0.6a$ and $0.8a$ for air--hole and InGaP features, respectively.}
\label{fig:L_simulation_data_gain_efficiency}
\end{figure}

Figure \ref{fig:L_simulation_data_gain_efficiency} shows the dependence of threshold gain and total external differential efficiency (sum of efficiencies at bottom and top surfaces $z_\mathrm{out}=0-\delta$ and $z_\mathrm{out}=L_z+\delta$, respectively) on the lengths of the sides of the active area. With rising $L$, $g_\mathrm{th}$ decreases and $\eta_\mathrm{d}$ increases first rapidly but change slowly for $L>400~\mu$m. 
The smaller value of the efficiency of the InGaP PCSEL compared to the air--hole PCSEL at larger values of $L$ can be mainly attributed to the higher waveguide loss $\alpha_\mathrm{WG}$.

\begin{figure}[!t]
\centering
\includegraphics[width=3in]{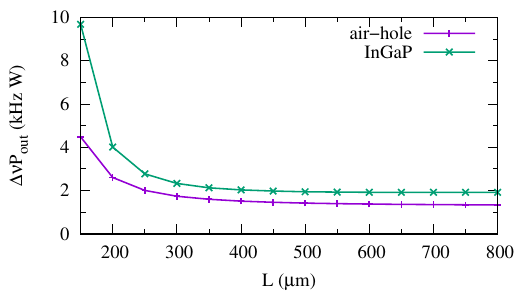}
\caption{Product of spectral linewidth and total output power of air--hole (violet) and InGaP (green) based PCSELs versus the lengths of the sides of the active area. Lengths of RIT legs  are $0.6a$ and $0.8a$ for air--hole and InGaP features, respectively.}
\label{fig:L_simulation_data_linewidth}
\end{figure}

\begin{figure}[!t]
\centering
\includegraphics[width=3in]{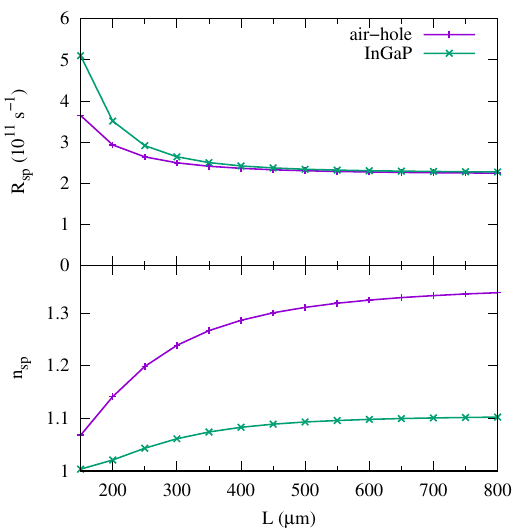}
\caption{Rate of spontaneous emission into the lasing mode (top) and population inversion factor (bottom) of air--hole (violet) and InGaP (green) based PCSELs versus the lengths of the sides of the active area. Lengths of RIT legs  are $0.6a$ and $0.8a$ for air--hole and InGaP features, respectively.}
\label{fig:L_simulation_data_Rsp_nsp}
\end{figure}

\begin{figure}[!t]
\centering
\includegraphics[width=3in]{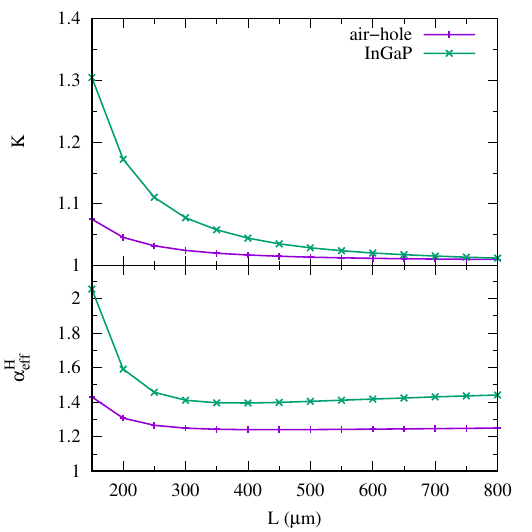}
\caption{Petermann factor (top) and effective Henry factor (bottom) of air--hole (violet) and InGaP (green) based PCSELs versus the lengths of the sides of the active area. Lengths of RIT legs  are $0.6a$ and $0.8a$ for air--hole and InGaP, respectively, features.}
\label{fig:L_simulation_data_Petermann_Henry}
\end{figure}

The linewidth--power product and contributing quantities are  plotted in Figs. \ref{fig:L_simulation_data_linewidth} -- \ref{fig:L_simulation_data_Petermann_Henry}. The decrease of $\Delta\nu P_\mathrm{out}$ is mainly caused by the decrease of threshold gain $g_\mathrm{th}$ and the rate of spontaneous emission $R_\mathrm{sp}$ with rising $L$, albeit the inversion factor $n_\mathrm{sp}=r_\mathrm{sp}(N_\mathrm{th})/g_\mathrm{a}(N_\mathrm{th})$ increases with rising $L$, in particular for the air--hole PCSEL (see Fig. \ref{fig:L_simulation_data_Rsp_nsp}). 
The air--hole PCSEL is distinguished by a larger value of $n_\mathrm{sp}$ which is the reason that $R_\mathrm{sp}$ is similar for both PCSEL types despite the lower threshold carrier density of the air--hole PCSEL (see Tab. \ref{tab:results}).

The decrease of the Petermann factor and the effective Henry factor for smaller values of $L$ contribute to the decrease of $\Delta\nu P_\mathrm{out}$, too.
For both types of PCSELs, $\Delta\nu P_\mathrm{out}$ has a similar magnitude in the range between $10$ and $2$~kHz$\cdot$W depending on $L$. Note, however, that a larger current has to be injected into the InGaP PCSEL to obtain the same output power as an air--hole PCSEL because of the lower efficiency and higher threshold current (see Tab. \ref{tab:results}).

\begin{table}[!t]
\caption{Linewidth power products \label{tab:linewidth} around $1064$~nm}
\centering
$\begin{array}{|c|c|c|c|}
\hline
\text{laser type} & \Delta\nu P_\text{out} \text{(sim)} & \Delta\nu P_\text{out} \text{(exp)}& \text{reference} \\
                  & \text{(kHz$\cdot$W)} & \text{(kHz$\cdot$W)} & \\
\hline\hline
\text{DFB}	& 1.3 ... 3 &  3.3 & \cite{spiessberger2010narrow}\\  
\text{DBR}	&  0.044 &  0.36 & \cite{spiessberger2011dbr}\\  
\text{mECDL}	& 0.009 &  0.058 & \cite{wenzel2022monolithically}\\  
\hline
\end{array}$
\end{table}

In Table \ref{tab:linewidth} experimental linewidth--power products (3rd column) are summarized for reported state--of--the--art EELs emitting around $1064$~nm.
The lowest value of $\Delta\nu P_\mathrm{out}$ was achieved with a monolithically integrated external cavity diode laser (mECDL). 
Simulated values of $\Delta\nu P_\mathrm{out}$ using the same theoretical model and global parameters as presented here are given in column 2 of Tab. \ref{tab:linewidth}. 
For the DFB laser a range is given because of the dependence of the linewidth on the phase of the non-vanishing rear facet reflectivity
\cite{nguyen2012optimization}.
Although the trend is predicted correctly, the simulated values of $\Delta\nu P_\mathrm{out}$ are smaller than the measured ones, in particular for the DBR laser and the mECDL.
There are several reasons that could explain the deviation. 
First, the global parameters given in Tab. \ref{tab:global} could not be appropriate for these devices because of differing active layers. Second, there a additional contributions to the intrinsic linewidth besides spontaneous emission as noted in the Introduction, in particular above threshold.
Finally, the experimental values might be limited by the experimental setups so that they represent an upper limit.
From Table \ref{tab:linewidth}, it can be concluded that the linewidth--product of the PCSELs studied here is similar to that of the DFB laser.
However, the output power of a high--power DFB laser is of the order $0.1$~W so that the minimum intrinsic linewidth is of the order $10$~kHz. In contrast, a PCSEL could emit tens of Watt resulting in intrinsic linewidths in the sub-kHz range.

\section{Summary}
A general theory for the calculation of the intrinsic spectral linewidth of photonic--crystal surface--emitting lasers has been presented.  
The analysis is based by treating spontaneous emission as a classical Langevin-type of noise source and by expanding the solution of the coupled--wave equations in terms of the solutions of the spectral problem.
The general expression of the linewidth includes the effective linewidth enhancement factor and the longitudinal excess factor of spontaneous emission.

The theoretical framework is used to compare air--hole and all-semiconductor PCSELs regarding their performance.
Although the air--hole PCSEL has a lower threshold gain and a higher external efficiency, the difference diminishes for large active regions.
The calculated linewidth--power product is of the order of several kHz$\cdot$W, with the prospect of very narrow linewidths being sustainable to Watt-class emission levels, in a narrow beam.
Together with their high--power capabilities, PCSELs emitting at $1064$~nm are therefore suited, \emph{e.g.}, for optical communication in space or non--linear frequency conversion.

\section*{Acknowledgment}
\noindent This work was partially performed in the frame of the project \emph{PCSELence} (K487/2022)
funded by the German Leibniz Association.

{\appendices

\section{Coupling matrix C}
\label{sec:Cmatrix}

According to \eqref{eq:Ansatz}, the coupling matrix can be written as the sum of three matrices,
\begin{equation}
	\bm{\mathrm{C}} = \bm{\mathrm{C}}_\mathrm{1D}+ \bm{\mathrm{C}}_\mathrm{rad}+ \bm{\mathrm{C}}_\mathrm{2D}.	
\end{equation}
The matrix $\bm{\mathrm{C}}_\mathrm{1D}$ describes the coupling of the forward and backward propagating waves like in a conventional DFB laser \cite{streifer1975coupling}. 
The matrix $\bm{\mathrm{C}}_\mathrm{rad}$ originates from the coupling via the out--of--plane (vertical) radiation fields and results in self and mutual coupling coefficients similar as originating in higher ($>1$) order Bragg gratings \cite{Shams2000}.
The matrix $\bm{\mathrm{C}}_\mathrm{2D}$ is due to the 2D in--plane coupling of the higher--order partial waves \cite{Liang2011}.
Whereas the matrices $\bm{\mathrm{C}}_\mathrm{1D}$ and $\bm{\mathrm{C}}_\mathrm{2D}$ are Hermitian if only the real part of the permittivity is taken into account in their computation, $\bm{\mathrm{C}}_\mathrm{rad}$ is not Hermitian. In particular, the imaginary part of the diagonal of $\bm{\mathrm{C}}_\mathrm{rad}$ (self--coupling coefficients) yields the surface radiation loss. In what follows, the notation of \cite{Liang2012} is partially adopted and results presented in  \cite{Peng2011} are utilized.

The 1D coupling matrix is given by
\begin{equation}
	\bm{\mathrm{C}}_\mathrm{1D} =
	\begin{pmatrix}
		0 & \kappa_{2,0} & 0 & 0 \\
		\kappa_{-2,0} & 0 & 0 & 0 \\
		0 & 0 & 0 & \kappa_{0,2} \\
		0 & 0 & \kappa_{0,-2} & 0 
	\end{pmatrix}
\end{equation}	
with the coupling coefficents
\begin{equation}
\begin{aligned}
\label{eq:kappa}
	\kappa_{\pm 2,0} & = -\frac{k_{0}^{2}}{2 \beta_{0}}\int_{z_\mathrm{min}}^{z_\mathrm{max}}\xi_{\pm 2, 0}(z)\vert \Theta(z)\vert^{2}dz,\\
	\kappa_{0,\pm 2} & = -\frac{k_{0}^{2}}{2 \beta_{0}}\int_{z_\mathrm{min}}^{z_\mathrm{max}}\xi_{0, \pm 2}(z)\vert \Theta(z)\vert^{2}dz.
\end{aligned}
\end{equation}
The radiation coupling matrix reads
\begin{equation}
\label{eq:Crad}
	\bm{\mathrm{C}}_\mathrm{rad} =
	\begin{pmatrix}
		\zeta_{1,0}^{(1,0)} & \zeta_{1,0}^{(-1,0)} & 0 & 0 \\
		\zeta_{-1,0}^{(1,0)} & \zeta_{-1,0}^{(-1,0)} & 0 & 0 \\
		0 & 0 & \zeta_{0,1}^{(0,1)} & \zeta_{0,1}^{(0,-1)} \\
		0 & 0 & \zeta_{0,-1}^{(0,1)} & \zeta_{0,-1}^{(0,-1)} 
	\end{pmatrix}
\end{equation}
 with 
\begin{multline}
\label{eq:zeta}	
	\zeta_{p,q}^{(r,s)} = -\frac{k_{0}^{4}}{2\beta_{0}}\int_{z_\mathrm{min}}^{z_\mathrm{max}}\xi_{p,q}(z)\\
	\times\int_{z_\mathrm{min}}^{z_\mathrm{max}}\xi_{-r,-s}(z')G(z,z')\Theta(z')dz'\Theta^{*}(z)dz
\end{multline}	
where $G(z,z')$ is the solution of \eqref{eq:Greeneq}.
It can be shown that the relations 
\begin{equation}
\zeta_{-1,0}^{(-1,0)}=\zeta_{1,0}^{(1,0)} ~\text{and}~
\zeta_{0,-1}^{(0,-1)}=\zeta_{0,1}^{(0,1)}
\end{equation}
hold.
The 2D coupling matrix is given by
\begin{equation}
	\bm{\mathrm{C}}_\mathrm{2D} =
	\begin{pmatrix}
		\chi_{y, 1,0}^{(1,0)}  & \chi_{y, 1,0}^{(-1,0)}  & \chi_{y, 1,0}^{(0,1)}  & \chi_{y, 1,0}^{(0,-1)} \\
		\chi_{y, -1,0}^{(1,0)} & \chi_{y, -1,0}^{(-1,0)} & \chi_{y, -1,0}^{(0,1)} & \chi_{y, -1,0}^{(0,-1)} \\
		\chi_{x, 0,1}^{(1,0)}  & \chi_{x, 0,1}^{(-1,0)}  & \chi_{x, 0,1}^{(0,1)}  & \chi_{x, 0,1}^{(0,-1)} \\
		\chi_{x, 0,-1}^{(1,0)} & \chi_{x, 0,-1}^{(-1,0)} & \chi_{x, 0,-1}^{(0,1)} & \chi_{x, 0,-1}^{(0,-1)}
	\end{pmatrix}
\end{equation}
with
\begin{equation}	
	\chi_{j, p,q}^{(r,s)} = -\frac{k_{0}^{2}}{2\beta_{0}}\sum_{\sqrt{m^2+n^2}>1}\varsigma_{j,m,n}^{(r,s)(p,q)},\ j=x,y,
\end{equation}	
\begin{multline}
	\begin{pmatrix}
		\varsigma_{x,m,n}^{(1,0)} & \varsigma_{x,m,n}^{(-1,0)} & \varsigma_{x,m,n}^{(0,1)} & \varsigma_{x,m,n}^{(0,-1)}\\
		\varsigma_{y,m,n}^{(1,0)} & \varsigma_{y,m,n}^{(-1,0)} & \varsigma_{y,m,n}^{(0,1)} & \varsigma_{y,m,n}^{(0,-1)}
	\end{pmatrix}^{(p,q)}\\ 
	=
	\frac{1}{m^{2}+n^{2}}
		\begin{pmatrix}
		n & m\\
		-m & n
	\end{pmatrix}	\\
	\times\begin{pmatrix}
		-m\mu_{m,n}^{(1,0)} & -m\mu_{m,n}^{(-1,0)} & n\mu_{m,n}^{(0,1)} & n\mu_{m,n}^{(0,-1)}\\
		 n\nu_{m,n}^{(1,0)} &  n\nu_{m,n}^{(-1,0)} & m\nu_{m,n}^{(0,1)} & m\nu_{m,n}^{(0,-1)}
	\end{pmatrix}^{(p,q)},
\end{multline}
\begin{multline}
\label{eq:mu}
\mu_{m,n}^{(r,s)(p,q)}
=k_{0}^{2}\int_{z_\mathrm{min}}^{z_\mathrm{max}}\xi_{p-m,q-n}(z)\\
\times\int_{z_\mathrm{min}}^{z_\mathrm{max}}\xi_{m-r,n-s}(z')G_{m,n}(z,z')\Theta(z')dz'\Theta^{*}(z)dz,
\end{multline}
\begin{multline}
\label{eq:nu}
\nu_{m,n}^{(r,s)(p,q)}
=-\int_{z_\mathrm{min}}^{z_\mathrm{max}}\frac{1}{\overline{\varepsilon}_\mathrm{s}(z)}\xi_{p-m,q-n}(z)\\
\times\xi_{m-r,n-s}(z)\vert \Theta(z)\vert^{2}dz.
\end{multline}
The following relations hold:
\begin{equation}
\begin{aligned}
\chi_{y,~1,0}^{~(~1,0)} & = \chi_{y,-1,~0}^{~(-1,~0)}, &  \chi_{x,~0,~1}^{~(~0,~1)} & = \chi_{x,0,-1}^{~(0,-1)},\\   
\chi_{y,-1,0}^{~(~0,1)} & = \chi_{x,~0,-1}^{~(~1,~0)}, &  \chi_{y,-1,~0}^{~(~0,-1)} & = \chi_{x,0,~1}^{~(1,~0)},\\    
\chi_{x,~0,1}^{~(-1,0)} & = \chi_{y,~1,~0}^{~(~0,-1)}, &  \chi_{x,~0,-1}^{~(-1,0)} & = \chi_{y,1,~0}^{~(0,~1)}.     
\end{aligned}
\end{equation}
Note that, if $\bm{\mathrm{C}}_\mathrm{1D}$ and $\bm{\mathrm{C}}_\mathrm{2D}$ are Hermitian (as it is the case here) and if $A(x,y)=A(y,x)$ (reflection symmetry) hold, further relations between the elements of the coupling matrices can be derived, \emph{c.f.} Ref. \cite{Inoue2022}.

The Greens's functions $G_{m,n}$ fulfill \cite{Liang2011}
\begin{multline}
\label{eq:Greeneq2}
\frac{d^2G_{m,n}(z,z')}{dz^2}+\left[k_0^2\overline{\varepsilon}_\mathrm{s}(z)-\beta^2(m^2+n^2)\right]\\
\times G_{m,n}(z,z')=-\delta(z-z')
\end{multline}
supplemented with homogeneous Robin boundary conditions.
Eqs. \eqref{eq:Greeneq} and \eqref{eq:Greeneq2} are  solved by a transfer matrix method  taking into account that the derivative of the Green's function jumps by one at $z=z'$. 
Integrals \eqref{eq:kappa} and, especially, \eqref{eq:zeta}, \eqref{eq:mu}, \eqref{eq:nu} are evaluated using analytic formulas for integrals of exponential functions, which compose $\Theta(z)$, $G(z,z')$, and $G_{m,n}(z,z')$.
Note that for $m^2+n^2\gg 1$ special care must be taken to avoid numerical overflow and underflow.
For more details, see \cite{Mindaugas2024} and \cite{Mindaugas2023}.

\section{Near and far field distributions}
\label{sec:NF_FF}

The intensity distribution of the near field is given by
\begin{equation}
|E_\mathrm{NF}|^2(x,y,t)=|E_{x,\mathrm{out}}(x,y,t)|^2+|E_{y,\mathrm{out}}(x,y,t)|^2
\end{equation}
where $E_{x|y,\mathrm{out}}(x,y,t)=\Delta E_{x|y}(x,y,z_\mathrm{out},t)$.
The far field distribution can be obtained in non-paraxial approximation by expressing the field in the outer region $z=z_\mathrm{out}$ in terms of a 2D Fourier integral, introducing spherical coordinates (radial distance $r$, azimuthal angle $\phi$, polar angle $\theta$), and employing the stationary phase method for large values of $r$ \cite{Kressel}. 
The result is \cite{Piprek}
\begin{equation}
\label{eq:ff1}
\vec{E}_{\text{FF}}(r,\phi,\theta,t) = \frac{e^{-ikr}}{ik^2r}k_z
\,\vec{e}(k_x,k_y,t)
\end{equation}
where 
\begin{equation}
\begin{aligned}
e_{x}(k_x,k_y,t)&=\frac{1}{2\pi}\iint_{-\infty}^{+\infty}E_{x,\mathrm{out}}(x,y,t)e^{i(k_xx + k_yy)}\,dxdy\\
e_{y}(k_x,k_y,t)&=\frac{1}{2\pi}\iint_{-\infty}^{+\infty}E_{y,\mathrm{out}}(x,y,t)e^{i(k_xx + k_yy)}\,dxdy\\
e_z(k_x,k_y,t)  &= -\frac{k_xe_x(k_x,k_y,t)+k_ye_y(k_x,k_y,t)}{k_z},
\end{aligned}
\end{equation}
\begin{equation}
\begin{aligned}
k_x&=k\sin{\theta}\cos{\phi}=k\frac{x}{r}\\
k_y&=k\sin{\theta}\sin{\phi}=k\frac{y}{r}\\
k_z&=\sqrt{k^2-k_x^2-k_y^2}=k\cos{\theta}=k\frac{z}{r},
\end{aligned}
\end{equation}
and $k=k_0n_\mathrm{out}$ with $n_\mathrm{out}$ being the index of the outer region.
The intensity distribution of the far field at a fixed distance $r$ is
\begin{multline}
|\vec{E}_\mathrm{FF}|^2(\phi,\theta)\propto(k_x^2+k_z^2)\langle|e_{x}(k_x,k_y,t)|^2\rangle\\
+(k_y^2+k_z^2)\langle|e_{y}(k_x,k_y,t)|^2\rangle
\end{multline}
omitting unnecessary prefactors.
Note that a two--dimensional plot of the far field is usually done in terms of the deflection angles
\begin{equation}
\begin{aligned}
\theta_x&=\mathrm{atan}\left(\frac{x}{z}\right)=\mathrm{atan}\Big[\tan(\theta)\cos(\phi)\Big]\\
\theta_y&=\mathrm{atan}\left(\frac{y}{z}\right)=\mathrm{atan}\Big[\tan(\theta)\sin(\phi)\Big].
\end{aligned}
\end{equation}
The deflection angles equal the polar angle, $\theta_x=\theta$ or $\theta_y=\theta$, for $\phi=0$ or $\phi=\pi/2$, respectively.

\section{Adjoint operator and mode orthogonality}
\label{sec:adjoint_ortho}

The spectral problem can be written as
\begin{equation}
\label{eq:spec}
\bm{\mathrm{L}}\vec{\Phi}\stackrel{\mathrm{def}}{=}v_\mathrm{g}\left[- \Delta\beta  + \bm{\mathrm{C}} - i\bm{\mathrm{D}}\right]\vec{\Phi} = \Omega  \vec{\Phi}
\end{equation}
where $\bm{\mathrm{L}}$ is a linear operator,
\begin{equation}
\vec{\Phi}(x,y)=\big(\Phi_u^+(x,y)~\Phi_u^-(x,y)~\Phi_v^+(x,y)~\Phi_v^-(x,y)\big)^\mathrm{T},
\end{equation} 
\begin{equation}
\label{eq:Dop}
\bm{\mathrm{D}}=\begin{pmatrix}
- \partial_x & 0 & 0 & 0\\
0 & + \partial_x & 0 & 0\\
0 & 0 & - \partial_y & 0\\
0 & 0 & 0 & + \partial_y
\end{pmatrix}
\end{equation}
is a diagonal first--order differential operator,
$^{T}$ denotes the transpose, and $\vec{\Phi}$ satisfies the boundary
conditions
\begin{equation}
\label{eq:bc1}
 \Phi_u^+(0,y)=\Phi_u^-(L,y)=\Phi_v^+(x,0)=\Phi_v^-(x,L)=0.
\end{equation}
An adjoint operator $\bm{\mathrm{L}}^\dagger$ can be defined via the integral relation
\begin{equation}
\label{eq:integralrelation}
\begin{aligned}
 &\iint \vec{\Psi}^T(x,y)\bm{\mathrm{L}}(x,y)\vec{\Phi}(x,y)\,dxdy\\
=&\iint \left[\bm{\mathrm{L}}^{\dagger}(x,y)\vec{\Psi}(x,y)\right]^\mathrm{T}\vec{\Phi}(x,y)\,dxdy
\end{aligned}
\end{equation}
where the spatial integration is made over the square $[0,L]\times[0,L]$.
The adjoint operator $\bm{\mathrm{L}}^{\dagger}$ is obtained by multiplying \eqref{eq:spec} from the left with
\begin{equation}
\vec{\Psi}^{\mathrm{T}}(x,y)=\big(\Psi_u^+(x,y)~\Psi_u^-(x,y)~\Psi_v^+(x,y)~\Psi_v^-(x,y)\big),
\end{equation} 
integrating and transposing,
\begin{multline}
\left[\iint \vec{\Psi}^\mathrm{T}\left[-\Delta \beta+\bm{\mathrm{C}}\right]\vec{\Phi}\,dxdy\right]^T\\
=\iint \vec{\Phi}^\mathrm{T}\left[-\Delta \beta+\bm{\mathrm{C}}^T\right]\vec{\Psi}\,dxdy
\end{multline}
and
\begin{equation}
\begin{aligned}
-i\left[\iint \vec{\Psi}^\mathrm{T}\bm{\mathrm{D}}\vec{\Phi}\,dxdy\right]^T
&=-i\iint (\bm{\mathrm{D}}\vec{\Phi})^\mathrm{T}\vec{\Psi}\,dxdy\\ 
&=-i\iint \vec{\Phi}^\mathrm{T}\overleftarrow{\bm{\mathrm{D}}}\vec{\Psi}\,dxdy\\
&=i\iint \vec{\Phi}^\mathrm{T}\bm{\mathrm{D}}\vec{\Psi}\,dxdy\\
&\hspace{5mm}+\underbrace{\text{boundary term}}_{\text{to be zero}}
\end{aligned}
\end{equation}
where the arrow directed to the left means that the differential operator $\bm{\mathrm{D}}^\mathrm{T}$ acts on $\vec{\Phi}^\mathrm{T}$.
Therefore, the adjoint operator is given by
\begin{equation}
\label{eq:adjoint}
\bm{\mathrm{L}}^{\dagger}\vec{\Psi}=v_\mathrm{g}\left[- \Delta\beta  + \bm{\mathrm{C}}^T + i\bm{\mathrm{D}}\right]\vec{\Psi}
\end{equation}
once $\vec{\Psi}$ fulfills the boundary conditions 
\begin{equation}
\label{eq:bcadj}
\Psi_u^+(L,y)=\Psi_u^-(0,y)=\Psi_v^+(x,L)=\Psi_v^-(x,0)=0
\end{equation}
obtained from the requirement of the vanishing boundary term
\begin{equation*}
\begin{aligned}
&\quad\int \left[\Phi_u^+\Psi_u^+|_{x=L}-\Phi_u^+\Psi_u^+|_{x=0}\right]\,dy\\
&-\int \left[\Phi_u^-\Psi_u^-|_{x=L}-\Phi_u^-\Psi_u^-|_{x=0}\right]\,dy\\
&+ \int \left[\Phi_v^+\Psi_v^+|_{y=L}-\Phi_v^+\Psi_v^+|_{y=0}\right]\,dx\\
&-\int \left[\Phi_v^-\Psi_v^-|_{y=L}-\Phi_v^-\Psi_v^-|_{y=0}\right]\,dx=0,
\end{aligned}
\end{equation*}
originating from the partial integration.

From definition \eqref{eq:integralrelation}, it follows that both operators $\bm{\mathrm{L}}$ and $\bm{\mathrm{L}}^{\dagger}$ share the same eigenvalue spectrum, \emph{i.e.}
\begin{equation}
\label{eq:adjointspectrum}
\bm{\mathrm{L}}^{\dagger}\vec{\Psi}=\Omega\vec{\Psi}.
\end{equation}
Considering an eigenfunction $\vec{\Phi}_\mu$ and an adjoint eigenfunction $\vec{\Psi}_\nu$ with eigenvalues $\Omega_\mu \ne \Omega_\nu$, respectively, it can be easily shown that the orthogonality equation
\begin{equation}
\label{eq:orthogonality}
\iint \vec{\Psi}_\mu^\mathrm{T}\vec{\Phi}_\nu\,dxdy = 0 \quad\text{for}\quad{\mu \ne \nu }
\end{equation}
holds.
Instead to solve \eqref{eq:adjointspectrum},
the adjoint eigenfunction $\vec{\Psi}$ can be directly determined from $\vec{\Phi}$
because of the general form 
\begin{equation}
\label{eq:C2Dstruc}
	\bm{\mathrm{C}} =
	\begin{pmatrix}
		C_{11} & C_{12} & C_{13}  & C_{14}  \\[1ex]
		C_{21} & C_{11} & C_{41}  & C_{31}  \\[1ex]
		C_{31} & C_{14} & C_{33}  & C_{34}  \\[1ex]
		C_{41} & C_{13} & C_{43}  & C_{33} 
	\end{pmatrix}
\end{equation}
of the coupling matrix due to the structure of the matrices $\bm{\mathrm{C}}_\mathrm{1D}$, $\bm{\mathrm{C}}_\mathrm{rad}$, and $\bm{\mathrm{C}}_\mathrm{2D}$. 
Therefore, $\Psi_u^+=\Phi_u^-$, $\Psi_u^-=\Phi_u^+$, $\Psi_v^+=\Phi_v^-$, and $\Psi_v^-=\Phi_v^+$ hold.
Note, that these relations can be considered to be a consequence of time--reversal symmetry \cite{marani1995spontaneous}.
The orthogonality relation \eqref{eq:orthogonality} can be written as
\begin{equation}
\label{eq:orthogonality2}
\left(\vec{\Phi}_\mu,\vec{\Phi}_\nu\right)=0\quad\text{for}\quad{\mu \ne \nu}
\end{equation}
with the inner product
\begin{multline}
\label{eq:inner}
\left(\vec{\Phi}_\mu,\vec{\Phi}_\nu\right)
=\iint \Big(\Phi^-_{u,\mu}\Phi^+_{u,\nu}+\Phi^+_{u,\mu}\Phi^-_{u,\nu}\\
+\Phi^-_{v,\mu}\Phi^+_{v,\nu}+\Phi^+_{v,\mu}\Phi^-_{v,\nu}\Big)\,dxdy.
\end{multline}
Note that at an exceptional point $\left(\vec{\Phi}_\mu,\vec{\Phi}_\nu\right)=0$ for $\mu=\nu$ \cite{wenzel1996mechanisms,ozdemir2019parity}.
The orthogonality relation of edge--emitting lasers is recovered if $\Phi^-_{v}=\Phi^+_{v}=0$ is set.

\begin{IEEEbiographynophoto}{Hans Wenzel}
received the Diploma and PhD. degrees in physics from Humboldt-Universit{\"a}t zu Berlin, Germany, in 1986 and 1991, respectively. His thesis dealt with the electro-optical modeling of semiconductor lasers. From 1991 to 1994, he held a post-doctoral position with Humboldt-Universit{\"a}t zu Berlin, where he was involved in the development of mathematical models for the stationary and dynamic simulation of distributed feedback lasers. 1993 he worked as a guest scientist at Tele Denmark Research in Hørsholm, Denmark. In 1994, Hans Wenzel joined Ferdinand-Braun-Institut (FBH), Berlin, Germany. Since then he headed the development of high-power distributed feedback and distributed Bragg reflector lasers for continuous-wave and pulsed operation. He designed new epitaxial layer structures for high brightness lasers, photonic integrated circuits, and quantum communication applications. His main research interests include the analysis, modeling, and simulation of optoelectronic devices. 
\end{IEEEbiographynophoto}

\begin{IEEEbiographynophoto}{Eduard Kuhn}
earned his M.Sc. degree in physics in 2017 from the University of Bremen, Germany, and completed his Ph.D. in 2022 at the Chemnitz University of Technology, Germany, specializing in the simulation of the longitudinal mode dynamics in Fabry-Pérot laser diodes.
Since 2023, he has been employed at the Weierstrass Institute in Berlin, Germany, where his primary research focus currently lies in the modeling and simulation of optoelectronic devices.
\end{IEEEbiographynophoto}

\begin{IEEEbiographynophoto}{Ben King}
received the Master of Physics degree from Durham University, Durham, U.K, in 2016, and the Ph.D. degree from The University of Glasgow, Glasgow, U.K,. in 2021. His Ph.D. thesis concerned the design, fabrication, and characterisation of Gallium Arsenide based photonic crystal surface emitting lasers. Since 2021 he has been a postdoctoral researcher at the Ferdinand-Braun-Institut, Berlin, Germany, where he works on the development of novel high-brightness semiconductor lasers.
\end{IEEEbiographynophoto}

\begin{IEEEbiographynophoto}{Paul Crump}
(M'97--SM'07) received the B.A. (Hons.) degree in
natural science (physics) from Oxford University, Oxford, U.K., and the doctoral
degree in physics from Nottingham University, Nottingham, U.K., in 1992 and
1996, respectively. His thesis dealt with experimental studies of 2D electrical
transport effects in semiconductor heterostructures. From 1996 to 2001, he was
with Agilent Technologies Corporation, Ipswich, U.K., where he helped de-
velop high-performance single-mode InP-based laser diodes. In 2001, he joined
nLight Photonics Corporation, Vancouver, WA, USA, where he helped develop
high-power, high-efficiency GaAs, and InP-based broad-area diode lasers and
bars. In 2007, he joined the Ferdinand-Braun-Institut (FBH), Berlin, Germany, where he leads the High-Power
Diode Laser Lab, that develops efficient, high power large area diode lasers
and their modules.
\end{IEEEbiographynophoto}

\begin{IEEEbiographynophoto}{Mindaugas Radziunas}
received the Diploma in Mathematics in 1992 
from Moscow State University, and the Ph.D. degree in 1996 from the 
Vilnius University, Lithuania, where he was specialized in numerical 
methods for nonlinear evolution equations. Since 1997 he has worked 
with the Weierstrass Institute, Berlin, Germany. 
His current research interests concentrate 
on modeling of multi-section semiconductor laser devices and numerical 
analysis of the model equations. 
\end{IEEEbiographynophoto}

\vfill

\end{document}